\documentclass[10pt]{book}
\usepackage{array}
\usepackage[usenames]{color}

\usepackage{amsmath,amsthm,amsfonts,amssymb,eucal,eufrak}
\usepackage{graphicx}
\usepackage{enumerate}

\newtheorem{Conjecture}{Conjecture}
\newtheorem{Corollary}{Corollary}
\newtheorem{Theorem}{Theorem}

\newtheorem{Lemma}{Lemma}
\font\bbigsym=cmmi10 scaled \magstep4 \font\bbbigsym=cmmi10 scaled
\def\prodi{\mathop{\lower 2pt\hbox{\bbigsym\char'031}}}
\def\Prodi{\mathop{\lower 2pt\hbox{\bbbigsym\char'031}}}
\newcommand{\ignore}[1]{}

\begin{document}


\renewcommand{\baselinestretch}{1.2}

\markright{
\hbox{\footnotesize\rm Statistica Sinica (in press)}\hfill
}

\markboth{\hfill{\footnotesize\rm YAAKOV MALINOVSKY and  YOSEF
RINOTT} \hfill} {\hfill {\footnotesize\rm Ordered Random Effects}
\hfill}

\renewcommand{\thefootnote}{}
$\ $\par


\fontsize{10.95}{14pt plus.8pt minus .6pt}\selectfont \vspace{0.8pc}
\centerline{\large\bf Prediction of Ordered Random Effects in a
Simple Small Area Model}
 \vspace{.4cm}
 \centerline{Yaakov Malinovsky$^{*}$  and Yosef Rinott$^{*,**}$}
\vspace{.4cm}
 \centerline{\it $^*$The Hebrew University of Jerusalem and $^{**}$LUISS, Rome}
 \vspace{.55cm}
 \fontsize{9}{11.5pt
plus.8pt minus .6pt} \selectfont

\begin{quotation}
\noindent {\it Abstract:}  Prediction of a vector of ordered parameters or  part of it arises
naturally in the context of Small Area Estimation (SAE). For
example, one may want to estimate the parameters associated with the
top ten areas, the best or worst area, or a certain percentile. We use a simple SAE
model to show that estimation of ordered parameters by the
corresponding ordered estimates of each area separately does not
yield good results with respect to MSE.  Shrinkage-type predictors,
with an appropriate amount of shrinkage for the particular problem of ordered parameters,
are considerably better, and their performance is close to that of
the optimal predictors, which cannot in general be computed
explicitly.
\par

\vspace{9pt} \noindent {\it Key words and phrases:} Empirical Bayes
predictor, Shrinkage, Order statistics, Linear predictor.
\par
\end{quotation}\par


\fontsize{10.95}{14pt plus.8pt minus .6pt}\selectfont

\section{Introduction}
\setcounter{chapter}{1} \setcounter{equation}{0}
We study the
prediction of ordered random effects in a simple model, motivated by
Small Area Estimation (SAE), under a quadratic loss function. The
model is
\begin{equation}
\label{Model:main}
 y_{i}=\mu+u_i+e_i,\quad i=1,...,m,
\end{equation}
where $y_i$ is  observed, $\mu$ is an unknown constant,
$u_i\stackrel{{iid}}\sim \textit{F}(0,\sigma_u^2)$  and
$e_i\stackrel{{iid}}\sim \textit{G}(0,\sigma_e^2)$, and F and G are
general distributions with zero means and variances $\sigma_u^2$ and
$\sigma_e^2$.  Set ${\bf{y}}=(y_1,...,y_m),
{\bf{u}}=(u_1,...,u_m)$, and ${\bf{e}}=(e_1,...,e_m)$, and assume
that $\bf{u}$ and $\bf{e}$ are independent. Set $\theta_i=\mu+u_i$ and $\boldsymbol{\theta}
=(\theta_1,\ldots,\theta_m)$.
The purpose is to predict the ordered random variables $\theta_{(i)}$,
$\left(\theta_{(1)}\leq\theta_{(2)}\leq...\leq\theta_{(m)}\right)$ from the observed
$y$'s. In SAE the random effect $\theta_{i}$ represent the $i$th area parameter.

The above model is a special case of  the SAE model of Fay and
Herriot (1979) that was presented in the context of estimating
per capita income for small places (i.e., population less than
1,000) from the 1970 Census of Population and Housing. The original
Fay-Herriot model allows different $\mu_{i}$ of the form
$\mu_{i}=x_i'\beta$, where $x_i$ is a vector of covariates for area
$i$, $\beta$ is a vector of coefficients that are common to all
areas, $u_i$ is a random effect of area $i$, and
$\theta_i=\mu_i+u_i$, the value of interest in area $i$, is measured
with a sampling error $e_i$. The SAE literature is concerned with
the estimation of $\theta_i$; see, e.g., Rao (2003). However, it is
also natural to consider the ordered parameters ${\theta}_{(i)}$ if
one is interested in estimating jointly the best, second best,
median, or worst area's parameter, for example,
or in studying the best or worst $k$ areas. In these
cases, one is interested in many or all ranked parameters, and not
just a single $\theta_{(i)}$.

When we have more than one observation per area,
the model is known as   the Battese-Harter-Fuller model (1988),
see also Pfeffermann (2002),
which we again simplify as in (\ref{Model:main}) :
\begin{equation}
\label{Model:Fuller}
y_{ij}=\mu+u_i+e_{ij},\quad  j=1,\ldots,n,\,\,i=1,\ldots,m.
\end{equation}
Typically in SAE $m$ is large, while $n$ is small;
however, we consider both small and large $m$.
Taking area means, as justified by sufficiency, the latter model reduces to that of \eqref{Model:main} with $\sigma_e^2$ replaced by
$\sigma_e^2/n$. When the  $\sigma_e^2$ is unknown, it should be estimated.  A main idea in SAE is  to  borrow strength across the different areas in order to predict effects. This can be applied also to variance estimation when some of the areas have only one observation; however, this is beyond the scope of the present paper, and for simplicity we assume the same number of observations $n$ in each area.

To see the difference between predicting the unordered vector $\boldsymbol\theta$, and
the ordered vector ${\boldsymbol\theta}_{(\,)}=(\theta_{(1)},\ldots,\theta_{(m)})$,
consider estimating the maximum $\theta_{(m)}$, and  two natural
unbiased predictors of
$\theta_i$, $\widehat{\theta}_i=y_i$ or
$\widehat{\theta}_i = E(\theta_i| \bf y)$.
By Jensen's inequality $\widehat{ \theta}_{(m)}
:=\max_i\widehat{\theta}_i$ is an overestimate  in expectation  of $\theta_{(m)}$  in the
case of $\widehat{\theta}_i=y_i$, and  an underestimate if we use
$\widehat{\theta}_i = E(\theta_i| \bf y)$. Such biases increase in $m$, which  in SAE and
in many parts of this paper is taken to be large. Similar considerations hold for other ordered parameters.

With different loss functions, prediction of the ordered parameters appears in Wright, Stern and Cressie (2003), and
prediction  and ranking of  small area parameters appear in Shen and Louis (1998). Their Bayesian methods require heavy numerical calculations, and
are sensitive to the choice of priors; see Shen and Louis (2000).
\ignore{In this paper we suggest simple to compute estimates, which are quite robust, and which
appear to be quite efficient.
Some of Louis' results will be compared to ours.
}  

If $\mu$ and
$\sigma_u^2$ and/or $\sigma_e^2$ are known, we have a Bayesian model in \eqref{Model:main} or \eqref{Model:Fuller}, and under quadratic loss, the optimal predictors
of the ordered parameters would be of the form
${\widehat{\theta}\,}_{(i)}=E(\theta_{(i)}|\bf y)$, where the expectation
depends on $\mu$,
$\sigma_u^2$ and $\sigma_e^2$, (and the distribution $F$ and $G$).
If $\mu$ and
$\sigma_u^2$ and/or $\sigma_e^2$ are unknown, we adopt an Empirical Bayes approach, and estimate them from the data.
However, even in the normal case, analytical computation of $E(\theta_{(i)}|\bf y)$ seems
intractable for $m>2$, and even more so under other distributions. Numerical computations
could be done, and in fact,  this is the Bayesian approach taken in principle by Wright et al (2003) and Shen
and Louis (1998, 2000);
the precise quantities they compute
are different due to the fact that they use different loss functions.

In this paper we avoid such Bayesian calculations and present simple predictors whose performance
is close to optimal; furthermore, due to their simplicity, they
are more robust against model misspecification.

Our starting point is the following.
Consider the predictor $\widehat{\theta}_{i}=E(\theta_{i}|\bf y)$ of $\theta_i$; under the assumption that
$F$ and $G$ are normal, we have
$\widehat{\theta}_{i}=\widehat{\theta}_{i}({\mu,\gamma^*})=\gamma^*y_i+(1-\gamma^*)\mu,$
where $\gamma^*=\sigma_u^2/(\sigma_u^2+\sigma_e^2)$.
For unknown $\mu$,  we plug in the estimator $\widehat{\mu}=\overline{y}$, and obtain
 the shrinkage-type predictor $\widehat{\theta}_{i}(\widehat{\mu})=\gamma^*y_i+(1-\gamma^*)\overline{y}$, which is the  best linear
unbiased predictor of $\theta_i$ for any $F$ and $G$; see, e.g.,  Robinson (1991), Rao (2003). Here $\gamma^*$
determines the amount of shrinkage toward the mean. We discuss the required amount of shrinkage when the goal is to predict ${\theta}_{(i)}$ rather than ${\theta}_{i}$.

For the problem of predicting the unordered parameters,
Bayesian considerations as above, and Stein (1956) and the ensuing huge body of literature suggest
shrinkage predictors. In  view of the discussion on under and overestimation, it
 is not surprising that  for the present problem
of predicting the ordered parameters, shrinkage is also desirable, but
to a lesser extent.  In fact, it can be shown geometrically that
if the coordinates of the predicting vector $\boldsymbol{{\widehat{\theta}}}
=(\widehat\theta_1,\ldots,\widehat\theta_m)$ happen to have the right order, that is, the same
order as the coordinates of ${\boldsymbol{\theta}}$,
then the desirable shrinkage is the same for the two problems, but otherwise
it is smaller. The latter case happens with high probability for large $m$
when the parameters are not very different. In this paper we show that rather
satisfactory results can be obtained by simple
predictors of the type
$\gamma y_{(i)}+(1-\gamma)\overline{y}$, and study the optimal value of $\gamma$. In general
we have $\gamma^* \le \gamma$.  Specifically, for large $m$ ($m>25$, say), we propose the predictor
$\sqrt{\gamma^*} y_{(i)}+(1-\sqrt{\gamma^*})\overline{y}$, to be denoted later by
${\widehat{\theta}\,^{[2]}_{(i)}}(\sqrt{\gamma^*})$. This predictor is easy to compute when the variances are either known
or estimated, and performs well in comparison to Bayes predictors, and other numerically demanding predictors
that appear in the literature.

In most of this paper we consider some predictor  $\widehat{{\boldsymbol{\theta}}}=
(\widehat{\theta}_{1},\ldots,\widehat{\theta}_{m}$), take $\widehat{\theta}_{(i)}$ as a predictor
of ${\theta}_{(i)}$
with a loss function given by
\begin{equation}
\label{eq:Loss} L({\widehat{\boldsymbol{\theta}}_{(\,)}},
\boldsymbol{\theta}_{(\,)})=\sum_{i=1}^m\left(\widehat{\theta}_{(i)}-\theta_{(i)}\right)^2,
\end{equation}
and compare different predictors in terms of the (Bayesian) risk
$$r(H,{\widehat{\theta}}\,) = E\{L({\widehat{\boldsymbol{\theta}}_{(\,)}},
\boldsymbol{\theta}_{(\,)})\},$$
where $H=(F,G)$ and the expectation is over all random variables involved.
Note that by a simple rearrangement inequality we always have
$\sum_{i=1}^m\left(\widehat{\theta}_{(i)}-\theta_{(i)}\right)^2 \le
\sum_{i=1}^m\left(\widehat{\theta}_{i}-\theta_{i}\right)^2$.

We also briefly consider the individual
mean square error (MSE) of a predictor $\widehat{\theta}_{(i)}$,
defined to be
$MSE(\widehat{\theta}_{(i)})=E(\widehat{\theta}_{(i)}-\theta_{(i)})^2$.

Even in the case of $m=2$ this prediction problem is not trivial.
Blumenthal and Cohen (1968) consider the following model: given
independent observation $X_1, X_2$ with $X_i\sim
N(\theta_i,\tau^2)$, estimate $\theta_{(2)}=\max(\theta_1,
\theta_2)$. They present five different estimators for
$\theta_{(2)}$ and evaluate their biases and mean square errors.
Generalizing their method to more than two parameters appears to be
hard.

Finally we mention that Senn (2008), with reference to Dawid (1994) and others,
deals with a different but related problem of estimation of the parameter $\theta_{i^*}$
corresponding to  $i^*=\arg \max y_i$. In SAE, this is the parameter belonging to the population
having the largest sample outcome, while we consider estimation of $\theta_{(m)}$,
the parameter of the ``largest population" (and likewise
for other ordered parameters).
The difference
is important when $m$ is large, and the parameters vary significantly;
our interest is in the ordered parameters and not parameters
chosen by the data, as in the above references.

In Section \ref{sec:pred} we discuss the model and present several
predictors. We also give minimax results that provide some justification to
normality assumptions and linear prediction. Section \ref{sec:main} contains the main results on
properties and comparisons of the various predictors. The proposed
class of predictors contains a parameter $\gamma$. Some of the
results apply to the whole class, while others suggest a range where
the best value of $\gamma$ should be, and apply to $\gamma$ in this
range. We describe a conjecture about the optimal value of $\gamma$ when $m$ is large and
provide an approximation for the optimal value of $\gamma$ in the normal case.
The last part of Section \ref{sec:main} deals with a special case when $F$ and $G$ are normal
and $m=2$. In this part we get  tighter conclusions than in the general case.

In Section 3 we assumed that the variances in  \eqref{Model:main} are known.
Section 4 deals with the case of unknown variances and studies plug-in Empirical Bayes predictors by simulation.
In Sections 5 and 6 we study robustness of the proposed predictors against
certain misspecifications of the assumptions on the distributions, and compare to
other predictors from the literature.

The proofs of results concerning general $m$ are given in Section 7.
The rest of the proofs are given in  an on-line
Supplement at\\ \textbf{http://www.stat.sinica.edu.tw/statistica}.
In the Supplement we provide simulations for Conjectures 1 and 2, we compare
various predictors under the assumption of known variances, and when one of the variances is
unknown. Theorems 5 and 6 are also proved there.

\par

\section{Predictors}\label{sec:pred}
\setcounter{chapter}{2} \setcounter{equation}{0}
\subsection{Unordered parameters  }
\setcounter{chapter}{2}
\setcounter{equation}{0}
In Sections \ref{sec:pred} and \ref{sec:main} we assume that
$\sigma_u^2$ and $\sigma_e^2$ are known. Later they are assumed
unknown, and plug-in estimators are used.
 First  we review some known results for the unordered case of Model (\ref{Model:main}) and
 the standard  problem of predicting $\theta_i$, $i=1,...,m$. The best linear  predictor
 is of the form
${\bf a}^{t}{\bf y}+b$, with ${\bf a},b$ that minimize the mean square
error. It is easy to see that when $\mu$ is known, and recalling that
$\sigma_u^2$ and $\sigma_e^2$ are now also assumed to be known, the best
linear predictor of $\theta_i$, that is, the predictor that minimized $E(\widehat{\theta}_{i}-\theta_{i})^2$ and therefore
$r^*(H,{\widehat{\theta}}\,):=E\left(\sum_{i=1}^m(\widehat{\theta}_{i}-\theta_{i})^2\right)$
among linear
predictors, is
\begin{equation}\label{minmax}
\widehat{\theta}_{i}({\mu})=\gamma^*y_i+(1-\gamma^*)\mu,
\end{equation}
where $\gamma^*=\sigma_u^2/(\sigma_u^2+\sigma_e^2)$.
Note that the model (\ref{Model:main}) does not
assume normality, and that the  best linear  predictor is unbiased,
that is, $E (\widehat{\theta}_{i}-{\theta}_{i})=0$.
When both distributions $F$ and $G$ are normal, the best linear
predictor  above is the best predictor (or Bayes predictor).

For unknown $\mu$,  the  best linear
unbiased predictor  (BLUP) of $\theta_i$ (see, for example, Robinson (1991), Rao (2003) and
references therein)  is
\begin{equation}
\label{E:EBl}
\widehat{\theta}_{i}(\widehat{\mu})=\gamma^*y_i+(1-\gamma^*)\overline{y},
\end{equation}
where $\widehat{\mu}:=\overline{y}=\frac{1}{m}\sum_{i=1}^m y_i$. The BLUP property
means that the predictor \eqref{E:EBl} minimizes $E(\widehat{\theta}_{i}-\theta_{i})^2$ among linear unbiased
predictors for all $F$ and $G$ with
the prescribed variances.  These are
shrinkage predictors (with
shrinkage towards the mean). Such predictors appear also in Fay and Herriot (1979).
We see in Section \ref{sec:main}
that for the ordered parameters shrinkage is also required, but
in a smaller amount (see also Louis (1984) and Ghosh (1992) for such shrinkage, under different
loss functions), showing again that the related
problems of predicting the ordered and
unordered parameters, are not the same.

\subsubsection{A justification of normality and linearity}

Using the fact that
an equalizer Bayes rule is minimax,  Schwarz (1987) proved the following result,
which in some sense justifies both linear estimators and the assumption of
normality of $F$ and $G$.

\begin{Theorem}\label{th:Sch}
Consider (\ref{Model:main})  with $\mu$, $\sigma_u^2$, and
$\sigma_e^2$ all fixed and known, and the risk function $r^*(H,{\widehat{\theta}}\,)=E\left(\sum_{i=1}^m(\widehat{\theta}_{i}-\theta_{i})^2\right)$.
The predictor  $\delta_{0}=(\delta_{01},\ldots,\delta_{0m})$ of  $\boldsymbol\theta$
given by $\delta_{0i}=\gamma^{*}y_i+(1-\gamma^{*})\mu,\,\,\,\,i=1,...,m$,
is minimax and the normal strategy  for $H=(F, G)$
is  least favorable.
\end{Theorem}

The next result is closely related to the previous one, and justifies linearity when
$\mu$ is unknown, which is the case we consider. It can easily be extended
to the original Fay-Herriot model with $\mu_i=x'_i\beta$.

\begin{Theorem}
Under the assumptions of Theorem \ref{th:Sch}, but with  unknown $\mu$, the predictor defined by $\delta_{0i}=\gamma^{*}y_i+(1-\gamma^{*})\overline{y},\,\,\,\,i=1,...,m$, is
minimax among all linear unbiased predictors of  $\boldsymbol\theta$.

\begin{proof}
Let $\cal H$ denote the class of pairs of distributions $(F,G)$ having the given variances.
Note  that $r^*(H,\delta_{0})$ depends only on the fixed variances, and therefore   for $H \in \cal H$,
$r^*(H,\delta_{0})$ is constant, say $v$. Let $\cal L$ denote the class of linear unbiased predictors.

We know that $\delta_{0}=(\delta_{01},\ldots,\delta_{0m})$ is BLUP.
We have
\begin{align*}
&\overline{V}=\inf_{\delta \in {\cal L}}\sup_{H \in {\cal H}}r^*(H,\delta)\leq \sup_{H \in {\cal H}}r^*(H,\delta_{0})=r^*(H_0,\delta_{0})=v,\\
&\underline{V}=\sup_{H \in {\cal H}}\inf_{\delta \in {\cal L}}r^*(H,\delta)\geq \inf_{\delta \in {\cal L}}r^*(H_{0},\delta)=r^*(H_{0},\delta_{0})
=v,
\end{align*}
for any $H_0 \in  \cal H$, where the penultimate equality holds by the BLUPness of $\delta_{0}$.
Since clearly $\overline{V} \ge \underline{V}$,
it follows that $\inf_{\delta \in {\cal L}}\sup_{H \in {\cal H}}r(H,\delta)= \sup_{H \in {\cal H}}r(H,\delta_{0})$,
so that $\delta_{0}$ is minimax among predictors in $\cal L$ as required.
\end{proof}
\end{Theorem}

\par
\subsection{Ordered parameters}\label{ordpar}
Let $\vartheta_{(i)}(\mu)=E_{{\mu}}(\theta_{(i)}|\bf y)$, the best
predictor of $\theta_{(i)}$ when $\mu$ is known, and consider its
empirical or plug-in version when $\mu$ is unknown:
 $E_{\widehat{\mu}}(\theta_{(i)}|{\bf y})=\vartheta_{(i)}(\widehat{\mu})$,
where   $\widehat{\mu}=\overline{y}$.

We  consider three predictors:
\begin{equation}
{\widehat{\theta}\,^{[1]}_{(i)}} =y_{(i)},\qquad
{\widehat{\theta}\,^{[2]}_{(i)}}(\gamma)=\gamma
y_{(i)}+(1-\gamma)\overline{y},\qquad
{\widehat{\theta}\,^{[3]}_{(i)}}=E_{\widehat{\mu}}(\theta_{(i)}|\bf
y),
\end{equation}
where $y_{(1)}\leq...\leq y_{(m)}$  denote the order statistics of
$y_1,...,y_m$.

Set ${\boldsymbol{\widehat{\theta}}\,^{[k]}_{(\,)}}=
\left({\widehat{\theta}\,^{[k]}_{(1)}},\ldots,{\widehat{\theta}\,^{[k]}_{(m)}}\right)$
for $k=1,2,3$. The predictors in the class
$\boldsymbol{\widehat{\theta}}\,^{[2]}_{(\,)}(\gamma)$ are analogous to the
best linear  predictor for the unordered case, but as we shall see,
the value of $\gamma$ has to be reconsidered, and
$\boldsymbol{\widehat{\theta}}\,^{[3]}_{(\,)}$ is the  empirical best predictor
in the ordered case (the best predictor with $\mu$ replaced by
$\overline{y}$). The latter predictor cannot in general be computed explicitly
for $m>2$, and some of our results are aimed at showing that it can
be efficiently replaced by
$\boldsymbol{\widehat{\theta}}\,^{[2]}_{(\,)}(\gamma)$ with an appropriate
choice of $\gamma$ for the ordered case at hand. Thus $\boldsymbol{\widehat{\theta}}\,^{[3]}_{(\,)}$,
the empirical best predictor, will be used as a yardstick to which other predictors are compared.

\section{Main results, known variances }\label{sec:main}
\setcounter{chapter}{3} \setcounter{equation}{0}

\subsection{General distributions F and G and general m}
\setcounter{chapter}{3}
\setcounter{equation}{0} 
The proofs of the results of this subsection are given in the Appendix.

 The first few results show that  shrinkage-type predictors in the
 class $\boldsymbol{\widehat{\theta}}\,^{[2]}_{(\,)}(\gamma)$ perform better
 than the predictor $\boldsymbol{\widehat{\theta}}\,^{[1]}_{(\,)}$. Refined calculations
 of the range of the optimal $\gamma$ allow us to understand the
amount of shrinkage required for the ordered parameters case.
\begin{Theorem}
\label{Th:big} Consider (\ref{Model:main}) with the
loss function (\ref{eq:Loss}) and
$\gamma^*=\sigma_u^2/(\sigma_u^2+\sigma_e^2)$. If
\begin{equation}
\label{Con:1}
\frac{m}{m-1}(2\sqrt{\gamma^{*}}-1)-\frac{1}{m-1}(2\gamma^{*}-1)\leq\,\,\,\gamma\,\,\,\leq1,
\end{equation}
then
\begin{equation}
\label{E:2vs1} E\{L(\boldsymbol{\widehat{\theta}}\,^{[2]}_{(\,)}(\gamma),
\boldsymbol{\theta}_{(\,)})\}\leq E\{L({\boldsymbol{\widehat{\theta}}\,^{[1]}_{(\,)}, \boldsymbol{\theta}_{(\,)})}\}.
\end{equation}
\end{Theorem}
Note that if  $\sigma^2_{u}\rightarrow 0$ then $\gamma^*
\rightarrow 0$ and the left-hand side of (\ref{Con:1}) tends to
$-1$. If $m\rightarrow\infty$, which is of interest in SAE, then the
left-hand side of (\ref{Con:1}) tends to $2\sqrt{\gamma^{*}}-1$. The
left-hand side of (\ref{Con:1}) is 1 when $\gamma^{*}=1$ and
increases in $\gamma^{*}$, hence  is bounded by 1.

By verifying the condition for the left-hand side of (\ref{Con:1})
to be nonpositive, we obtain the following.
\begin{Corollary}
\label{Cl:c1} If
\begin{equation}\label{eq:up}
\gamma^{*}\leq\left(\frac{m-\sqrt{(m-1)^2+1}}{2}\right)^2,
\end{equation}
 then
\begin{equation}\label{eq:mas}
E\{L(\boldsymbol{\widehat{\theta}}\,^{[2]}_{(\,)}(\gamma),
\boldsymbol{\theta}_{(\,)})\}\leq E\{L({\boldsymbol{\widehat{\theta}}\,^{[1]}_{(\,)}, \boldsymbol{\theta}_{(\,)})}\}
\end{equation}
for all $\gamma, 0\leq\gamma\leq1$.

\end{Corollary}

Note that asymptotically \eqref{eq:up} becomes
$\gamma^{*}\leq\frac{1}{4}$, since $\lim_{m\rightarrow\infty}
\left(\frac{m-\sqrt{(m-1)^2+1}}{2}\right)^2=\frac{1}{4}$.
Condition (\ref{eq:up}) is sufficient and may not be necessary.
But, (\ref{eq:mas}) does not hold without a suitable condition on $\gamma^*$; for
example, if $m=100$, the upper bound of (\ref{eq:up}) is 0.2475.
For $\gamma^*=1/3$ and $\gamma=0.1$, a straightforward simulation
using normal variables shows that (\ref{eq:mas}) does not hold.

For $\gamma=\gamma^{*}$,  (\ref{Con:1})
holds if and only if   $\gamma^{*}\leq (m-1)^2/(m+1)^2$  (see Appendix). From this
result we obtain the following.
\begin{Corollary}
\label{Cl:c2} If
\begin{equation}
\label{Con:2}
 \gamma^{*}\leq (m-1)^2/(m+1)^2,
\end{equation}
$\text{then}\quad E\{L(\boldsymbol{\widehat{\theta}}\,^{[2]}_{(\,)}(\gamma),
\boldsymbol{\theta}_{(\,)})\}\leq E\{L({\boldsymbol{\widehat{\theta}}\,^{[1]}_{(\,)}, \boldsymbol{\theta}_{(\,)})}\}
\,\,\, \text{ for all} \quad \gamma,\,
\gamma^{*}\leq\gamma\leq1.$
\end{Corollary}
Asymptotically, Corollary \ref{Cl:c2} holds for all $\gamma^*$ without
(\ref{Con:2}), because $\lim_{m\rightarrow\infty}\{
(m-1)^2/(m+1)^2\}=1$ and $0\leq\gamma^*\leq 1$ by definition. A
small simulation study indicates that Corollary \ref{Cl:c2} may hold
without the condition $\gamma^*\leq (m-1)^2/(m+1)^2$ for a large
variety of F and G. We can prove it
only for the extreme case $m=2$ and normal F and G;
see Theorem \ref{Th:small}  below. The range of $\gamma$'s for which
shrinkage improves the predictors, $\gamma^{*}\leq\gamma$, indicates
that, for the ordered problem, less shrinkage is required.

The following lemma is used in the proof of Theorem \ref{Th:big},
but may be of independent interest.
\begin{Lemma}
\label{Le:main} Under
(\ref{Model:main}),
$$m(\sigma_u^2+\mu^2)\leqslant
E\sum_{i=1}^m{\theta_{(i)}}{y}_{(i)}\leqslant
m[(\sigma_u^2+\mu^2)(\sigma_u^2+\sigma_e^2+\mu^2)]^{1/2}.$$
\end{Lemma}
For the predictors ${\widehat{\theta}\,^{[2]}_{(i)}}(\gamma)$, it is
natural to look for optimal or good values of $\gamma$.
\begin{Theorem} \label{Th:MinPoint} Under
(\ref{Model:main}), let $\gamma^{\,o}$ be the optimal choice of
$\gamma$ for the predictor
${\widehat{\theta}\,^{[2]}_{(i)}}(\gamma)$ in the sense of
minimizing $E\{L(\boldsymbol{\widehat{\theta}}\,^{[2]}_{(\,)}(\gamma),
\boldsymbol{\theta}_{(\,)})\}$.
Then
\begin{equation}
\label{E:OptR}
\gamma^{\,o} \in \left[\gamma^{*},
\frac{m}{m-1}\sqrt{\gamma^*}-\frac{1}{m-1}\gamma^*\right].
\end{equation}
\end{Theorem}
As $m\rightarrow\infty$, the above range for the optimal  $\gamma$
becomes $\left[\gamma^{*}, \sqrt{\gamma^*}\right].$

\begin{Conjecture}
\label{Conj:C1} The optimal $\gamma$ in the sense of Theorem
\ref{Th:MinPoint} satisfies
$\lim_{m\rightarrow\infty}\gamma^{\,o}=\sqrt{\gamma^{*}}$.
\end{Conjecture}
Simulations that justify  Conjecture \ref{Conj:C1} are given in the Supplement.

For $m>25$ or so, which is common in SAE, we recommend using the predictor
$\boldsymbol{\widehat{\theta}}\,^{[2]}_{(\,)}(\gamma)$ with $\gamma=\sqrt{\gamma^*}$. Numerous simulations
suggest that the latter choice, or the choice of $\gamma = \gamma^{\,o}$, yield essentially
the same results. We emphasize that the predictor $\boldsymbol{\widehat{\theta}}\,^{[2]}_{(\,)}(\sqrt{\gamma^*})$
is very easy to compute.
The case of $m \le 25$ is discussed next.

\subsection{An approximation for $\gamma^{\,o}$ in the normal case}
\label{sec:approx}
For practical computation of $\gamma^{\,o}$ for $m \le 25$ or so, we
propose the following approach that we have implemented in the normal
case. (For $m>25$, $\sqrt{\gamma^*}$ provides an excellent
approximation to $\gamma^{\,o}$, see simulation results and
Conjecture \ref{Conj:C1}). In view of Theorem \ref{Th:MinPoint} we
consider the approximation formula
\begin{equation}\label{eq:approxgamma0}
{\gamma^{\,o}}\approx\alpha \gamma^* +
(1-\alpha)u(m,\gamma^*),
\end{equation}
with
$u(m,\gamma^*)=\frac{m}{m-1}\sqrt{\gamma^*}-\frac{1}{m-1}\gamma^*$,
and $\alpha$ depending on $m$ and $\gamma^*$. For fixed $\gamma^*$,
and for each $m$ satisfying $2 \le m \le 30$ we compute
$E\{L(\boldsymbol{\widehat{\theta}}\,^{[2]}_{(\,)}(\gamma),
\boldsymbol{\theta}_{(\,)})\}$ by  simulations
and find the minimizer $\gamma^{\,o}$ by an exhaustive search. We
then define $\alpha_{m,\gamma^*}$ to be the solution of
(\ref{eq:approxgamma0}).
For fixed   $\gamma^*$ we carry out polynomial regression of the computed values of
$\alpha_{m,\gamma^*}$ on the explanatory variable $m$ in the  range $2 \le m
\le 30$;  this is repeated for an array of values of $\gamma^*$.
It turns out that an excellent approximation is obtained
when $\alpha_{m,\gamma^*}=\alpha_m$ is taken to be only a function of $m$ for a
large range of values of $\gamma^*$. We therefore combine the
different regressions for the different values of $\gamma^*$, and
obtain a polynomial approximation for $\alpha_m$. The numerical
calculations lead to the quadratic polynomial
$\alpha_m=0.8236-0.0573m+0.0012m^2$. Plugging it into
(\ref{eq:approxgamma0}) we obtain the approximation
$\widetilde{\gamma^{\,o}}=\alpha_m \gamma^* +
(1-\alpha_m)u(m,\gamma^*)$ for $\gamma^{\,o}$.

Numerical simulations show that in the range $2 \le m \le 25$ and
for all values of $\gamma^*$, the resulting
$\widetilde{\gamma^{\,o}}$ is indeed very close to $\gamma^{\,o}$.
In fact, for $m=2$ they may differ by about $10\%$, but for $m \ge 4$ they differ
 by about $1\%-2\%$. Using one or the other yields almost identical expected
losses.

\subsection{Normal distribution of F and G and m=2}
 When both F and G are normal and $m=2$, we  obtain tighter conclusions for the previous
results.
\begin{Theorem}
\label{Th:small} For (\ref{Model:main}) with $F$ and $G$
normal and $m=2$:
\begin{enumerate}
\item if \quad $0\leq\gamma^*\leq c \thickapprox0.4119$, then
$E\{L(\boldsymbol{\widehat{\theta}}\,^{[2]}_{(\,)}(\gamma),
\boldsymbol{\theta}_{(\,)})\}\leq E\{L({\boldsymbol{\widehat{\theta}}\,^{[1]}_{(\,)}, \boldsymbol{\theta}_{(\,)})}\}$ for all\quad
$0\leq\gamma\leq1$;
 \item for all $\gamma^*$ and $\gamma$ satisfying
$\gamma^{*}\leq\gamma\leq 1$,
$E\{L(\boldsymbol{\widehat{\theta}}\,^{[2]}_{(\,)}(\gamma),
\boldsymbol{\theta}_{(\,)})\}\leq E\{L({\boldsymbol{\widehat{\theta}}\,^{[1]}_{(\,)}, \boldsymbol{\theta}_{(\,)})}\}$;
\item the optimal
$\gamma$ for the predictor
${\widehat{\theta}\,^{[2]}_{(i)}}(\gamma)$ (i=1,2) in the sense
of minimizing $E\{L(\boldsymbol{\widehat{\theta}}\,^{[2]}_{(\,)}(\gamma),
\boldsymbol{\theta}_{(\,)})\}$
is
\begin{align}
\label{Opt:m2}
&\gamma^{\,o}=\gamma^*\left(4\psi(a)-1\right)+(1-\gamma^*)\frac{2}{\pi}\sqrt{\gamma^{*}(1-\gamma^{*})},
\end{align}
where $\psi(a)=\int_{0}^{\infty}t^2
\Phi\left({{a}}t\right)\varphi(t)dt$,
and $a=\sqrt{\frac{\gamma^{*}}{1-\gamma^{*}}}.$

\end{enumerate}
\end{Theorem}
Thus Part \textit{1} of Theorem \ref{Th:small} shows that we can replace
the condition  $\gamma^*\leq \left(\frac{2-\sqrt{2}}{2}\right)^2
\thickapprox 0.086$ of Corollary \ref{Cl:c1} by $\gamma^*\leq c,
\,\,c\thickapprox0.4119$;
Part \textit{2} shows that
(\ref{Con:2}) ($\gamma^*\leq 1/9$ for $m=2$)
 of Corollary \ref{Cl:c2} may be omitted;
Part \textit{3} of Theorem \ref{Th:small} gives an exact result
 rather than the range given by (\ref{E:OptR}).

\noindent\textbf{ Remark}.
The accurate definition of $c$ and its approximation are given
 in the proof of Theorem \ref{Th:small} in the Supplement. The
function $\psi(a)$ can be computed by Matlab:
$\text{double}(int(x^2*\text{normpdf}\,(x)*(\text{erf}\,(a*x/sqrt(2))+1)/2,x,0,\text{inf}\,))$.

The results given so far compare
${\widehat{\theta}_{(i)}}^{\,\,{[2]}}(\gamma)$ with
${\widehat{\theta}_{(i)}}^{\,\,{[1]}}$ in the sense of minimizing
expected loss. In the absence of an explicit  expression for
${\widehat{\theta}_{(i)}}^{\,\,{[3]}}$, it is  not easy to compare
it with other predictors analyticallly, but it is possible to do this
if  F and G are normal and $m=2$, and the
result is Theorem \ref{Th:smallBest}. For $m>2$ we provide
simulations.

It is obvious that the estimator
$\vartheta_{(i)}(\mu)=E_{{\mu}}(\theta_{(i)}|y)$ minimizes the MSE.
The point of Theorem \ref{Th:smallBest} is that  the unknown $\mu$
is replaced in $\vartheta_{(i)}(\mu)$ by its estimate $\overline{y}$
to obtain ${\widehat{\theta}}_{(i)}^{\,\,{[3]}}$.

\begin{Theorem}
\label{Th:smallBest} Consider (\ref{Model:main}) with $F$ and $G$
normal. Then for   $m=2$,
$E\{L({\boldsymbol{\widehat{\theta}}\,^{[3]}_{(\,)}, \boldsymbol{{\theta}}_{(\,)})}\}\leq
E\{L({\boldsymbol{\widehat{\theta}}\,^{[2]}_{(\,)}(\gamma^{\,*}), \boldsymbol{{\theta}}_{(\,)})}\}$.
\end{Theorem}
\begin{Conjecture} \label{conj:E} If $F$ and $G$ are normal and $m\geq2$, then
$E\{L({\boldsymbol{\widehat{\theta}}\,^{[3]}_{(\,)}, \boldsymbol{{\theta}}_{(\,)})}\}\leq
E\{L({\boldsymbol{\widehat{\theta}}\,^{[2]}_{(\,)}(\gamma^{\,o}), \boldsymbol{{\theta}}_{(\,)})}\}$.
\end{Conjecture}
Various simulations support this conjecture. Some of them are presented in the Supplement.
The simulations  show that  the predictor
$\boldsymbol{\widehat{\theta}}\,^{[2]}_{(\,)}(\gamma^{\,o})$
is worse than
$\boldsymbol{\widehat{\theta}}\,^{[3]}_{(\,)}$ in the sense of
$E\{L({\boldsymbol{\widehat{\theta}}}_{(\,)}, \boldsymbol{\theta}_{(\,)})\}$, as suggested by Conjecture 2. However, they are rather close, while the predictor
$\boldsymbol{\widehat{\theta}}\,^{[2]}_{(\,)}(\gamma^{\,*})$  is far
worse. This suggests    that the linear predictor
$\boldsymbol{\widehat{\theta}}\,^{[2]}_{(\,)}(\gamma^{\,o})$  can be
used without much loss. As mentioned above, for $m \ge 25$ or so, the predictor
$\boldsymbol{\widehat{\theta}}\,^{[2]}_{(\,)}(\sqrt{\gamma^{\,*}})$, which is easy
to calculate, is
as good as $\boldsymbol{\widehat{\theta}}\,^{[2]}_{(\,)}(\gamma^{\,o})$,
and the calculation of $\gamma^{\,o}$ can be avoided. See the Supplement.
\par

\section{Unknown variances}
\setcounter{chapter}{4} \setcounter{equation}{0}
Until now it was assumed that the variances are known. We now turn to the case of unknown variances.
This case will be studied by simulations, whose detailed description is given in
the Supplement.

We first make the common assumption in SAE that only
$\sigma_u^2$ is unknown, and later that both variances, $\sigma_u^2$
and $\sigma_e^2$ are unknown. We replace
each unknown variance by plugging-in its natural estimator. For the case that only
$\sigma_u^2$ is unknown, it is estimated by
\begin{align}
    \label{EVar:plugin}
    \widehat{\sigma}_u^2=\max\left(\frac{1}{m-1}\sum_{i=1}^m \left(y_{i}-\overline{y}\right)^2-\sigma_e^2,0\right).
    \end{align}
This approach cannot be expected to work for small values of
$m$. We emphasize again the interest in SAE is in large $m$'s.

The notation for the resulting estimates remains as
it was for the case of known variances. In this case, and in the case that both
variances are unknown (Figure 1 below), we use simulations to
compare  the risk $ E\{L({\boldsymbol{\widehat{\theta}}}_{(\,)}, \boldsymbol{\theta}_{(\,)})\}$
for the predictors $\boldsymbol{\widehat{\theta}}\,^{[3]}_{(\,)}$,\,
$\boldsymbol{\widehat{\theta}}\,^{[2]}_{(\,)}({\gamma^*})$,\,
$\boldsymbol{\widehat{\theta}}\,^{[2]}_{(\,)}(\sqrt{\gamma^*})$, and
$\boldsymbol{\widehat{\theta}}\,^{[2]}_{(\,)}(\gamma^{\,o})$, and
since in all simulations
the risks of the latter two predictors are almost identical, we present only
one of them. We also compare the performance of these predictors when only
the maximum is predicted.

\begin{center}
\includegraphics[height=1.40in]
{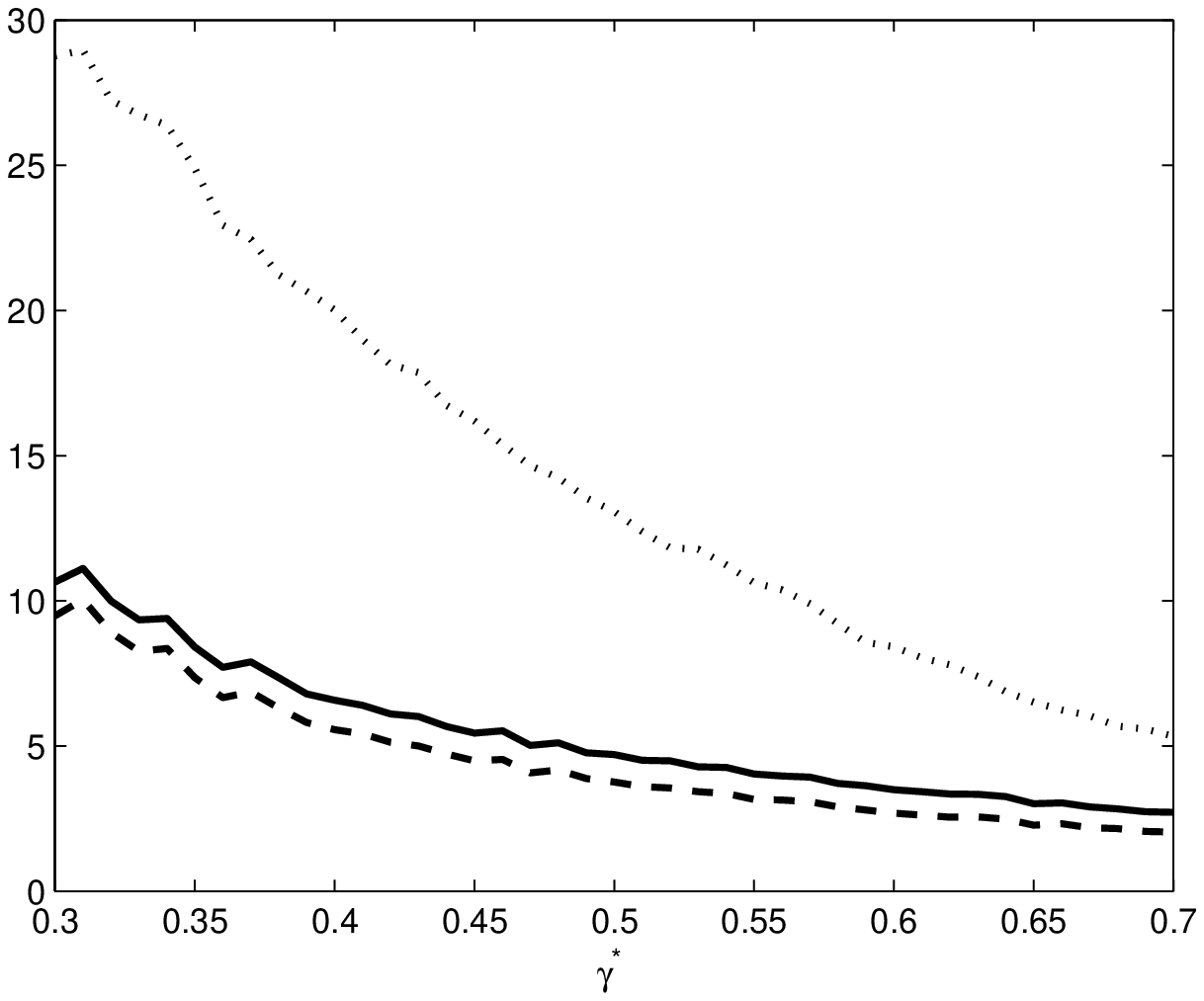}
\includegraphics[height=1.40in]
{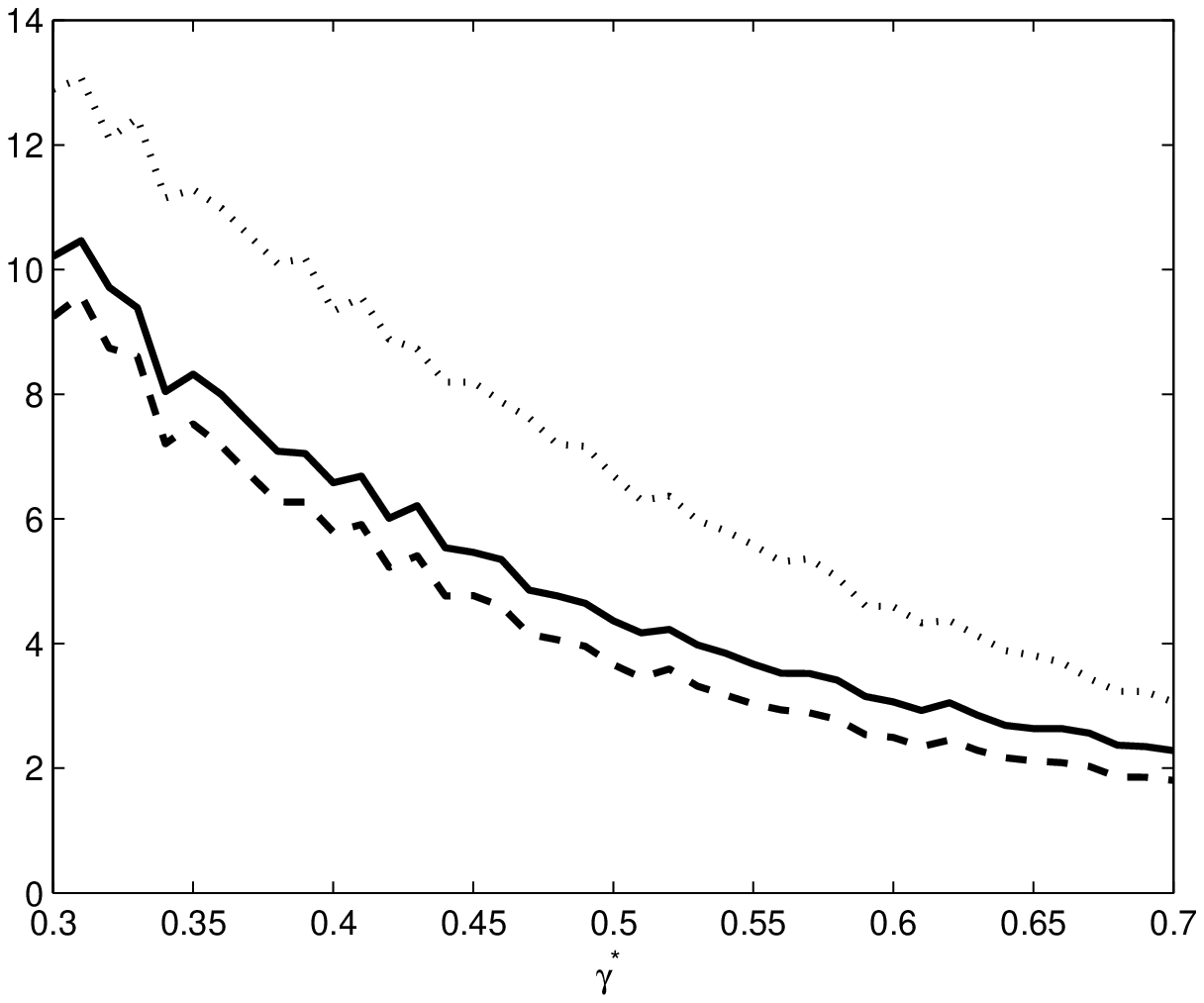}
\includegraphics[height=1.40in]
{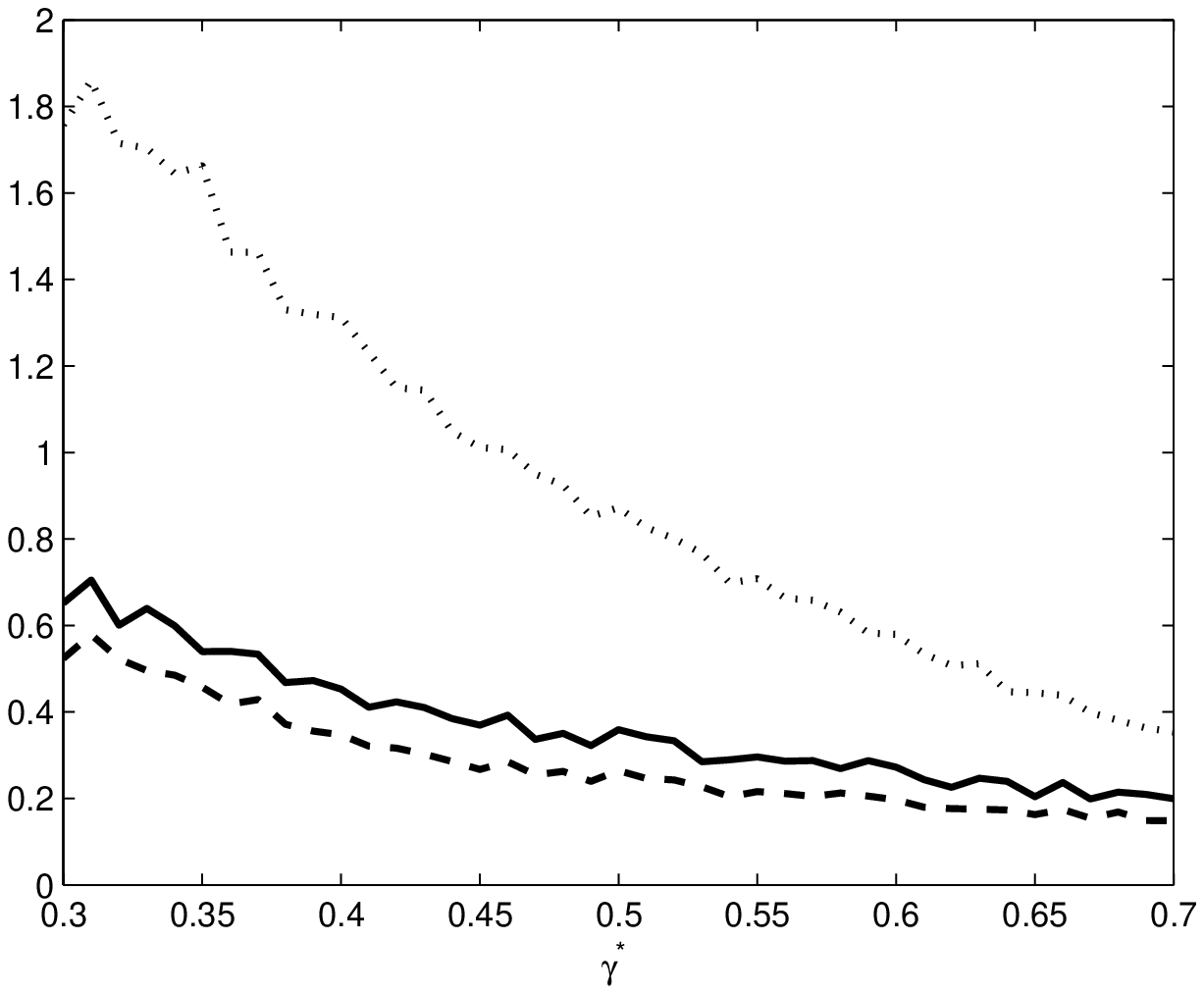}
\includegraphics[height=1.40in]
{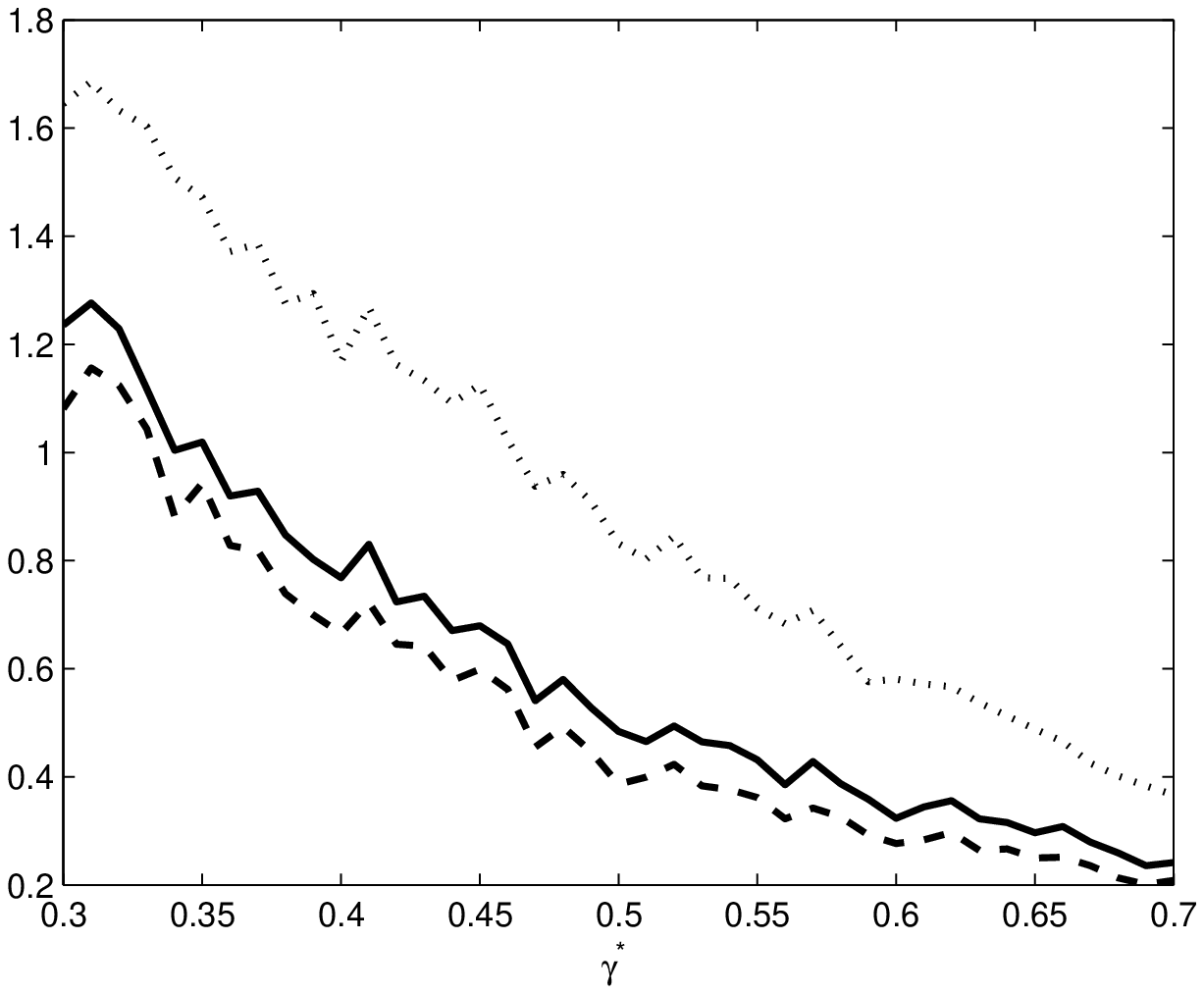}
\end{center}
\textbf{Figure 1:}

\begin{center}
\begin{itemize}
\item
 Comparison of $ E\{L({\boldsymbol{\widehat{\theta}}}_{(\,)}, \boldsymbol{\theta}_{(\,)})\}$ as a function of $\gamma^*,$ for the predictors
$\boldsymbol{\widehat{\theta}}\,^{[2]}_{(\,)}({\gamma^*})$,\,
$\boldsymbol{\widehat{\theta}}\,^{[2]}_{(\,)}(\sqrt{\gamma^*})$,\,
$\boldsymbol{\widehat{\theta}}\,^{[3]}_{(\,)}$ (dotted, solid, dashed lines),
where F and G are normal and\\ $m=100,n=15$ (upper left), $m=30,n=15$ (upper right)
\item
Comparison of the MSE of
${\widehat{\theta}_{(m)}}^{\,\,{[2]}}(\gamma^*),\,
{\widehat{\theta}_{(m)}}^{\,\,{[2]}}(\sqrt{\gamma^*}),\,
{\widehat{\theta}_{(m)}}^{\,\,{[3]}}$ (dotted, solid, dashed lines)
for predicting $\theta_{(m)}$, as a function of $\gamma^*$, where F and G
are normal and $m=100,n=15$ (bottom left), $m=30,n=15$ (bottom right)
\end{itemize}
\end{center}

In Figure 3S (given in the Supplement) only
$\sigma_u^2$ is estimated, and in Figure 1 both $\sigma_u^2$ and
$\sigma_e^2$ are estimated. The figures are rather similar.
The results should be compared to those
of Figure 2S (Supplement), where the variances are known. Clearly the less one
knows, the higher the loss. However, the simple shrinkage predictor
$\boldsymbol{\widehat{\theta}}\,^{[2]}_{(\,)}(\sqrt{\gamma^*})$  performed almost as well
as the best plug-in predictor $\boldsymbol{\widehat{\theta}}\,^{[3]}_{(\,)}$, and
much better than $\boldsymbol{\widehat{\theta}}\,^{[2]}_{(\,)}({\gamma^*})$. Thus again we conclude that for
the problem at hand, shrinkage estimators work, provided one uses
the right amount of shrinkage for the ordered parameters problem.

For the case of unknowns ${\sigma}_u^2$ and ${\sigma}_e^2$, consider the model
$$y_{ij}=\mu+u_i+e_{ij}; i=1,...,m ,j=1,...,n,$$ which is a
special case of the Nested Error Unit Level Regression model of
Battese, Harter and Fuller (1988). We apply our previous estimators,
replacing the variances by
$$\widehat{\sigma}_e^2=\frac{1}{m(n-1)}\sum_{i=1}^m\sum_{j=1}^{n}
 \left(y_{ij}-\overline{y}_{i.}\right)^2, \quad
\widehat{\sigma}_u^2=\max\big\{\frac{1}{mn-1}\sum_{i=1}^m\sum_{j=1}^{n}
 \left({y}_{ij}-\overline{y}_{..}\right)^2-\widehat{\sigma}_e^2,0\big\},$$
and set $\displaystyle \gamma^{*}=\frac{\hat\sigma_u^2}{\hat\sigma_u^2+\hat\sigma_e^2/n}$.
Simulation results are given Figure 1.

\section{Shrinkage type predictor
$\boldsymbol{\widehat{\theta}}\,^{[2]}_{(\,)}(\sqrt{\gamma^*})$ in the non normal case}
\setcounter{chapter}{5} \setcounter{equation}{0}
\label{Subsec:shrinkage}
We briefly consider non-normal $F$, whereas the error distribution $G$ remains normal.
We first take the double exponential distribution (Laplace distribution) for the random effects $u_i$, with density
$\frac{1}{2b}\exp\left(-\frac{|u_i|}{b}\right)$,
where $b=\frac{\sigma_{u}}{\sqrt{2}}$.
Direct calculations show that the density function of $\theta_i$ given $y_i$ is
\begin{align}
\label{CDis:Laplace}
&f_{\theta_{i}|y_{i}}\left(t|y\right)=
\begin{cases}
\displaystyle \frac{p_{1}(t)}{\int_{-\infty}^{\mu}p_{1}(t)dt+\int_{\mu}^{\infty}p_{2}(t)dt},&\text{if $t\leq \mu$}\\
\displaystyle \frac{p_{2}(t)}{\int_{-\infty}^{\mu}p_{1}(t)dt+\int_{\mu}^{\infty}p_{2}(t)dt},&\text{if $t>\mu$}
\end{cases}
\end{align}
where
$p_{i}(t)=exp\left(-{\left(t-\left(y+(-1)^{i+1}{\sigma_{e}^2 }{b^{-1}}\right)\right)^2}/{2\sigma_{e}^2}\right),\,\,\,
i=1,2$.

\begin{center}
\includegraphics[height=1.40in]{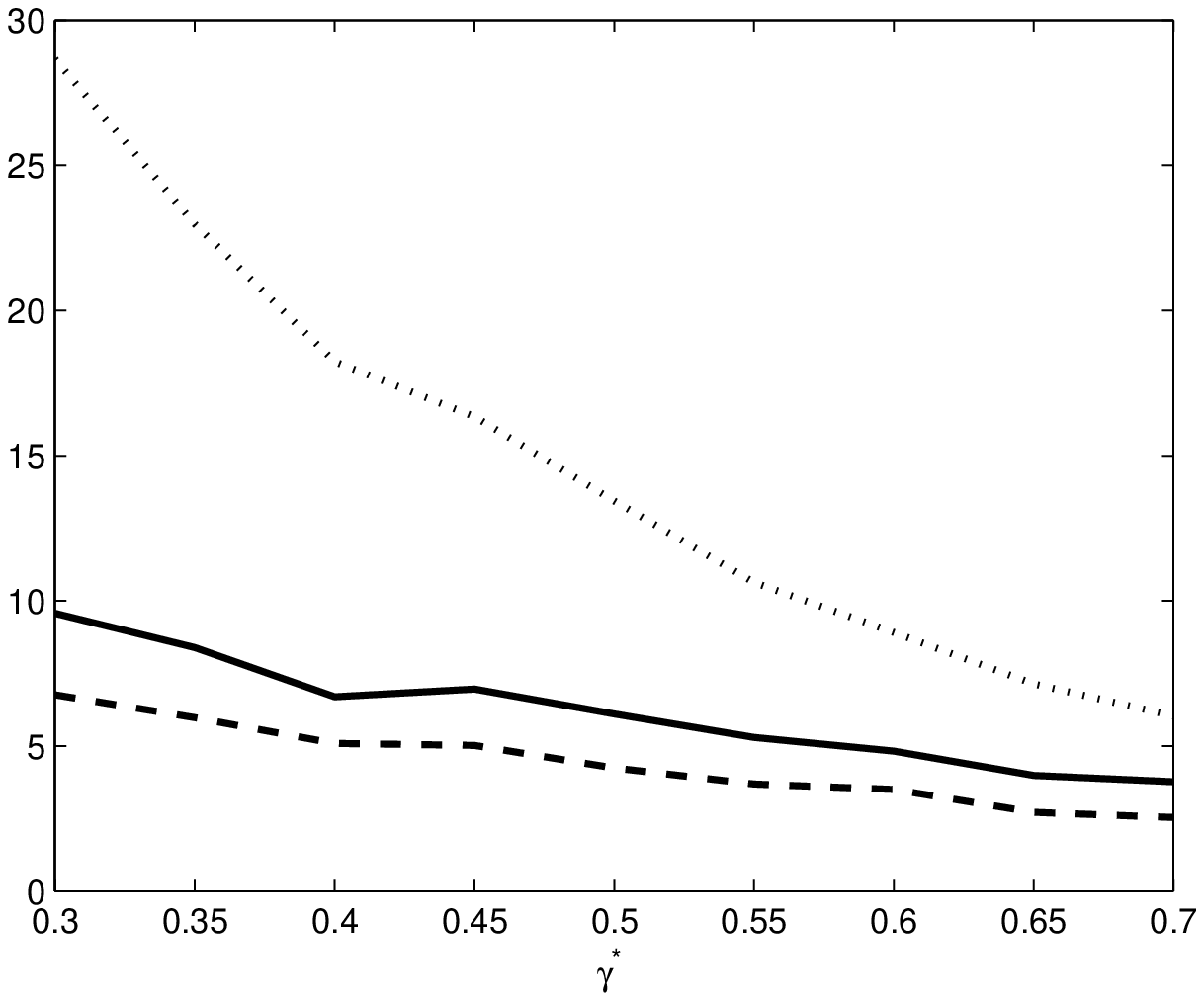}
\includegraphics[height=1.40in]{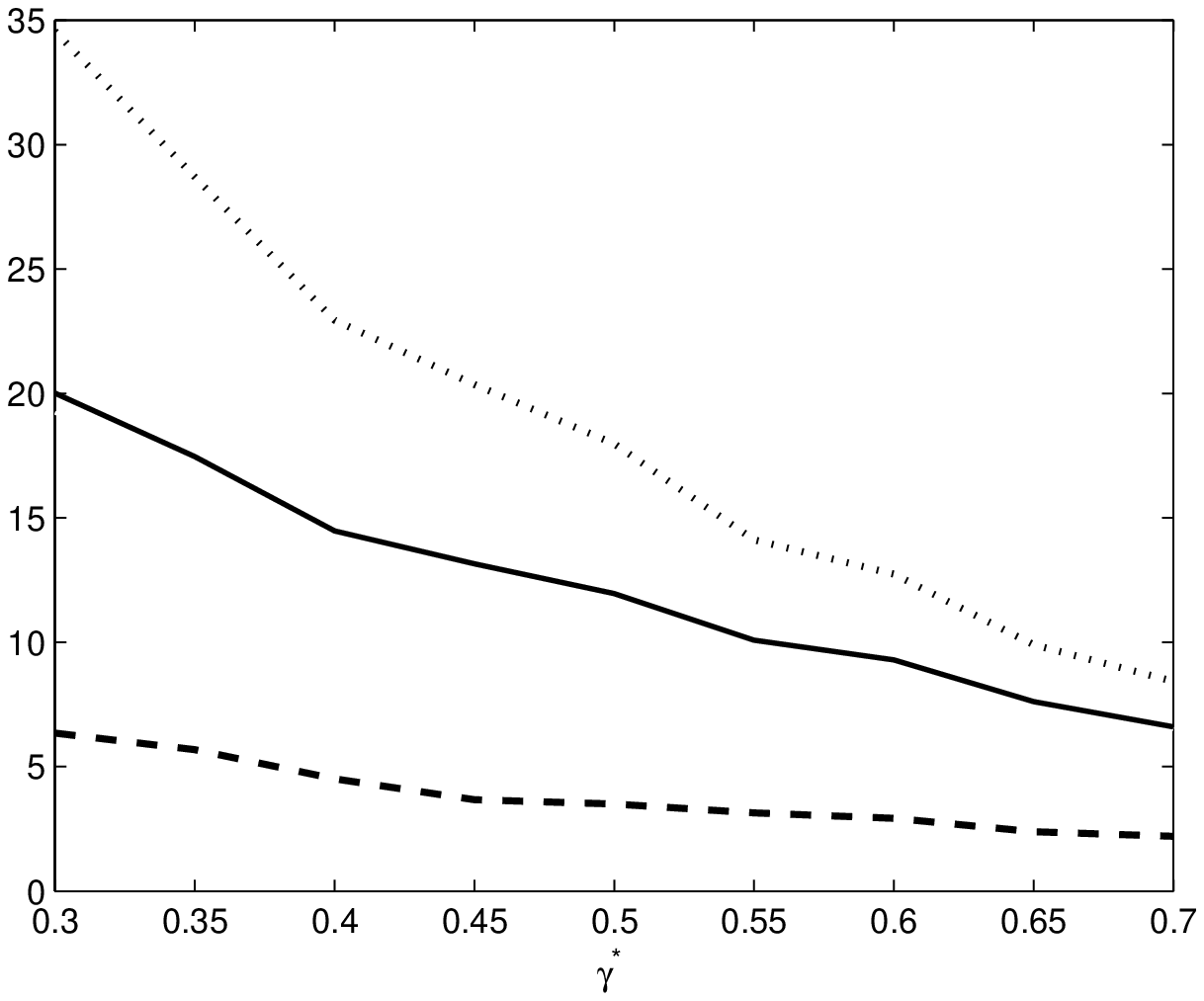}
\includegraphics[height=1.40in]{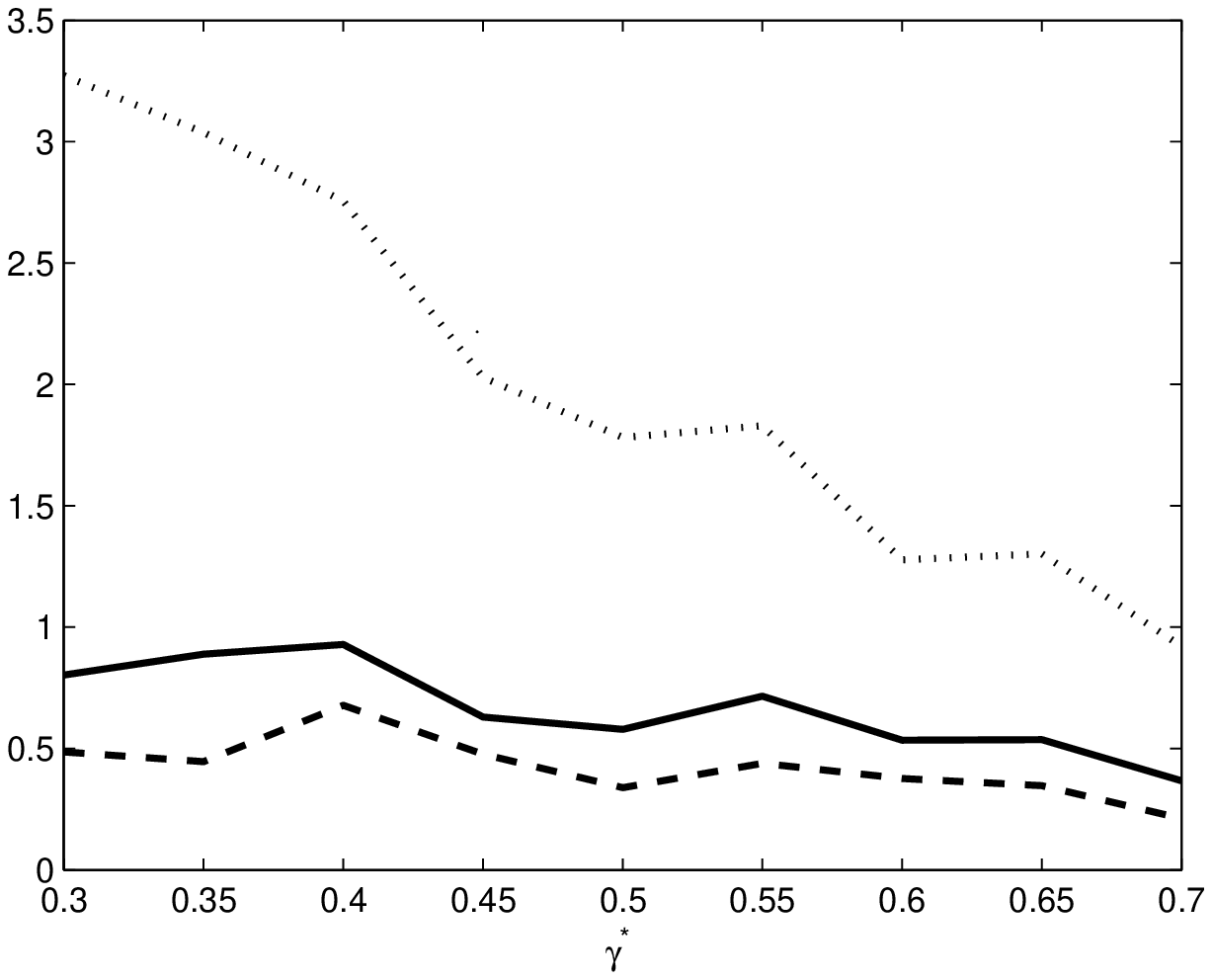}
\includegraphics[height=1.40in]{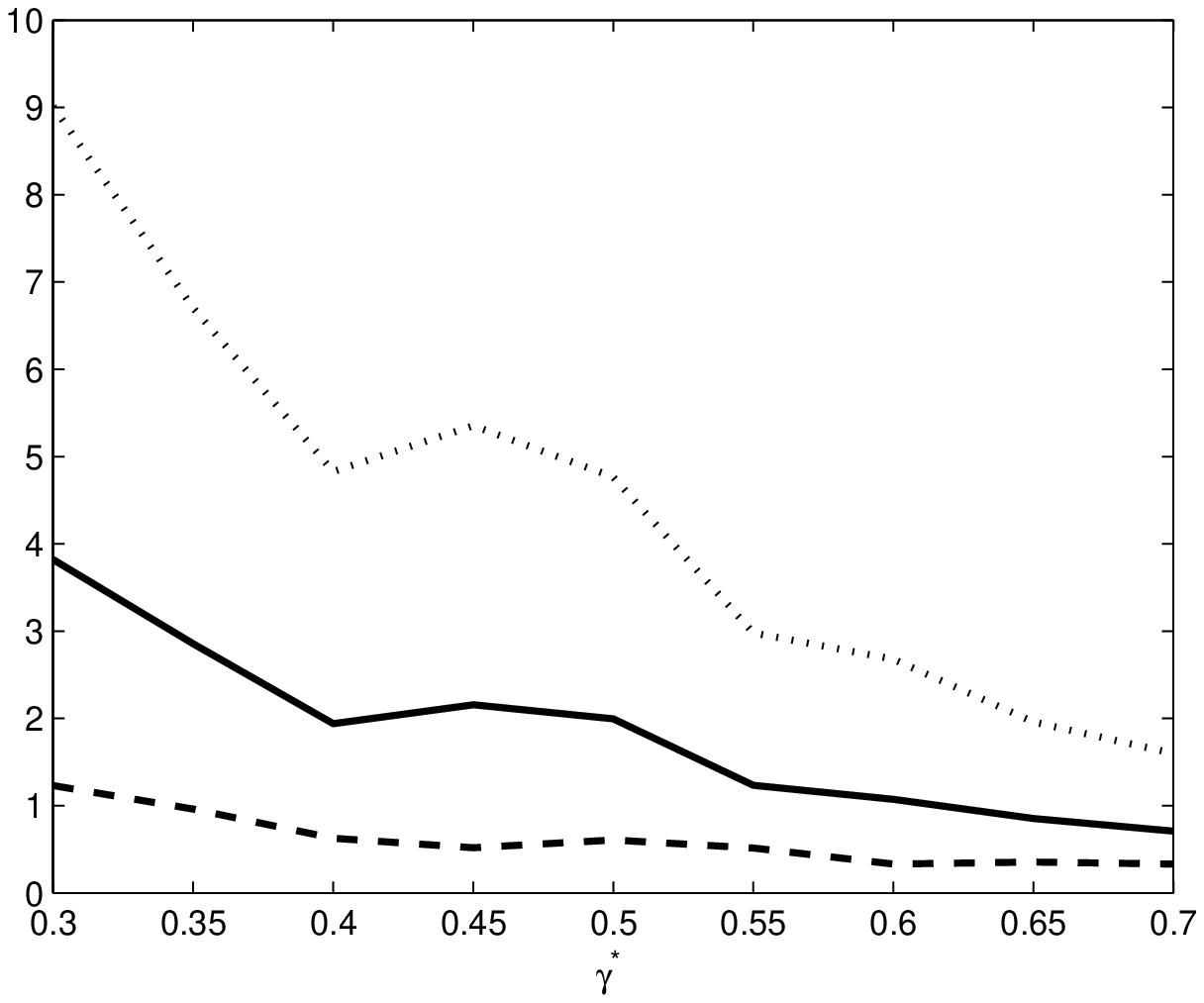}
\end{center}

\textbf{Figure 2:}

\begin{center}
\begin{itemize}
\item
 Comparison of $ E\{L({\boldsymbol{\widehat{\theta}}}_{(\,)}, \boldsymbol{\theta}_{(\,)})\}$ as a function of $\gamma^*,$ for the predictors
$\boldsymbol{\widehat{\theta}}\,^{[2]}_{(\,)}({\gamma^*})$,\,
$\boldsymbol{\widehat{\theta}}\,^{[2]}_{(\,)}(\gamma^{\,o})$,\,
$\boldsymbol{\widehat{\theta}}\,^{[3]}_{(\,)}$
(dotted, solid, dashed lines),
where F is the Laplace distribution and G is normal (upper left), and
where F is the Location exponential distribution and G is normal (upper right),
for $m=100$.
\item
 Comparison of the MSE of
${\widehat{\theta}_{(m)}}^{\,\,{[2]}}(\gamma^*),\,
{\widehat{\theta}_{(m)}}^{\,\,{[2]}}(\gamma^{\,o}),\,
{\widehat{\theta}_{(m)}}^{\,\,{[3]}}$ (dotted, solid, dashed lines)
for predicting $\theta_{(m)}$, as a function of $\gamma^*$,
where F is the Laplace distribution and G is normal (bottom left), and
where F is the Location exponential distribution and G is normal (bottom right),
for $m=100$.
\end{itemize}
\end{center}
We also take a location exponential distribution for the random effects , with density
$\frac{1}{b}\exp\left(-\frac{u_i-a}{b}\right)1_{(u_i\geq a)}$,
where $b=\sigma_u,\,\,a=-b$.
By direct calculation the density function of $\theta_i$ given $y_i$ is
\begin{align}
\label{CDis:LExp}
&f_{\theta_{i}|y_{i}}\left(t|y\right)=\displaystyle
\frac{exp\left(-{\left(t-\left(y-{\sigma_e^2}{\sigma_u^{-1}}\right)\right)^2}/
{2\sigma_e^2}\right)1_{(t\geq-\sigma_u+\mu)}}{\int_{-\sigma_u+\mu}^{\infty}
exp\left(-{\left(t-\left(y-{\sigma_e^2}{\sigma_u^{-1}}\right)\right)^2}/
{2\sigma_e^2}\right)dt}
\end{align}

The  simulations (Figure 2) were done as in Figure 2S (m=100), except that
for each value of $\gamma^*$ we ran 100 simulations and
generated 100 random variables
from $f_{\theta_{i}|y_{i}}\left(\cdot|\cdot\right)$, sorted them, and approximated ${\widehat{\theta}_{(i)}}^{\,\,{[3]}}.$

We can see in Figure 2 that for the symmetric but heavy-tailed Laplace distribution, our
shrinkage type predictor
$\boldsymbol{\widehat{\theta}}\,^{[2]}_{(\,)}({\gamma^o})$ (and the same
is true for $\boldsymbol{\widehat{\theta}}\,^{[2]}_{(\,)}({\sqrt{\gamma^*}})$\,) is
close to the empirical best predictor $\boldsymbol{\widehat{\theta}}\,^{[3]}_{(\,)}$,
but in the asymmetric case of the Location Exponential distribution, this does not happen.

\section{Robustness and comparison with Shen and Louis (1998)}
\setcounter{chapter}{6} \setcounter{equation}{0}

Shen and Louis (1998; henceforth SL) proposed predictors called ``Triple-goal estimates" for  random effects in two-stage hierarchical models.
Their method is in general not analytically
tractable, and requires numerical
calculations.  Moreover, being sensitive to Bayesian assumptions,   it is not robust (Shen and Louis (2000)).

The first stage of SL is
minimizing  $E\int\left\{A(t;{\bf y})-G_m(t)\right\}^2 dt$ with the constraint that $A$
is a discrete distribution with at most $m$ mass points, where $G_{m}(t)$ is the `empirical' distribution
function
$G_{m}(t)=\frac{1}{m}\sum_{i=1}^m I_{(\theta_{i}\leq t)}$.
They show that the solution  $A$ is the empirical distribution of  $\boldsymbol{\widehat{U}}=(\widehat{U}_1,...,\widehat{U}_m)$,
$\widehat{U}_{j}=\overline{G}_{m}^{\,-1}\left(\frac{2j-1}{2m}\right)$, where
$\overline{G}_{m}(t)=E\left(G_{m}(t)|\bf y\right)=\frac{1}{m}\sum_{i=1}^{m}P\left(\theta_{i}\leq t|y_{k}\right)$.
Therefore $\widehat{U}_{j}$ is a predictor of $\theta_{(j)}$,\,\,\,$j=1,...,m$.
The solution $\widehat{U}_{j}=\overline{G}_{m}^{\,-1}\left(\frac{2j-1}{2m}\right)$
depends on the posterior distributions of $\theta_1,....,\theta_m$ and requires estimation of unknown parameters
and a solution of nonlinear equations.
In order to compute  $\overline{G}_{m}(t)$ in our simulations, we compute
$P\left(\theta_{i}\leq t|y_{k}\right)$ using the plug-in (or moment) estimator $\overline{y}$ of $\mu$,
and \eqref{Model:main} with the assumption that $F$ and $G$ are normal,
and apply Matlab function $\mathbf{`fzero'}$  for the solution $t=\widehat{U}_{j}$ of the equations $\overline{G}_{m}(t)=\frac{2j-1}{2m}$.

For the purpose of checking robustness we generated data taking $F$ to be the Laplace distribution
or the asymmetric location exponential distribution, and
a normal $G$.
The  simulations were done as in Figure 2S (m=100), except that in the stage of prediction we ignored
the true distribution of the random effects and used the normal distribution.
Here we compared
$\boldsymbol{\widehat{\theta}}\,^{[2]}_{(\,)}(\sqrt{\gamma^*}),\,
\boldsymbol{\widehat{\theta}}\,^{[3]}_{(\,)}$,  and the predictor $\boldsymbol{\widehat{U}}$ based on SL.
Note that, unlike the estimators in SL, it is not necessary to know the distributions for the predictor
$\boldsymbol{\widehat{\theta}}\,^{[2]}_{(\,)}(\sqrt{\gamma^*})$.

\begin{center}
\includegraphics[height=1.40in]{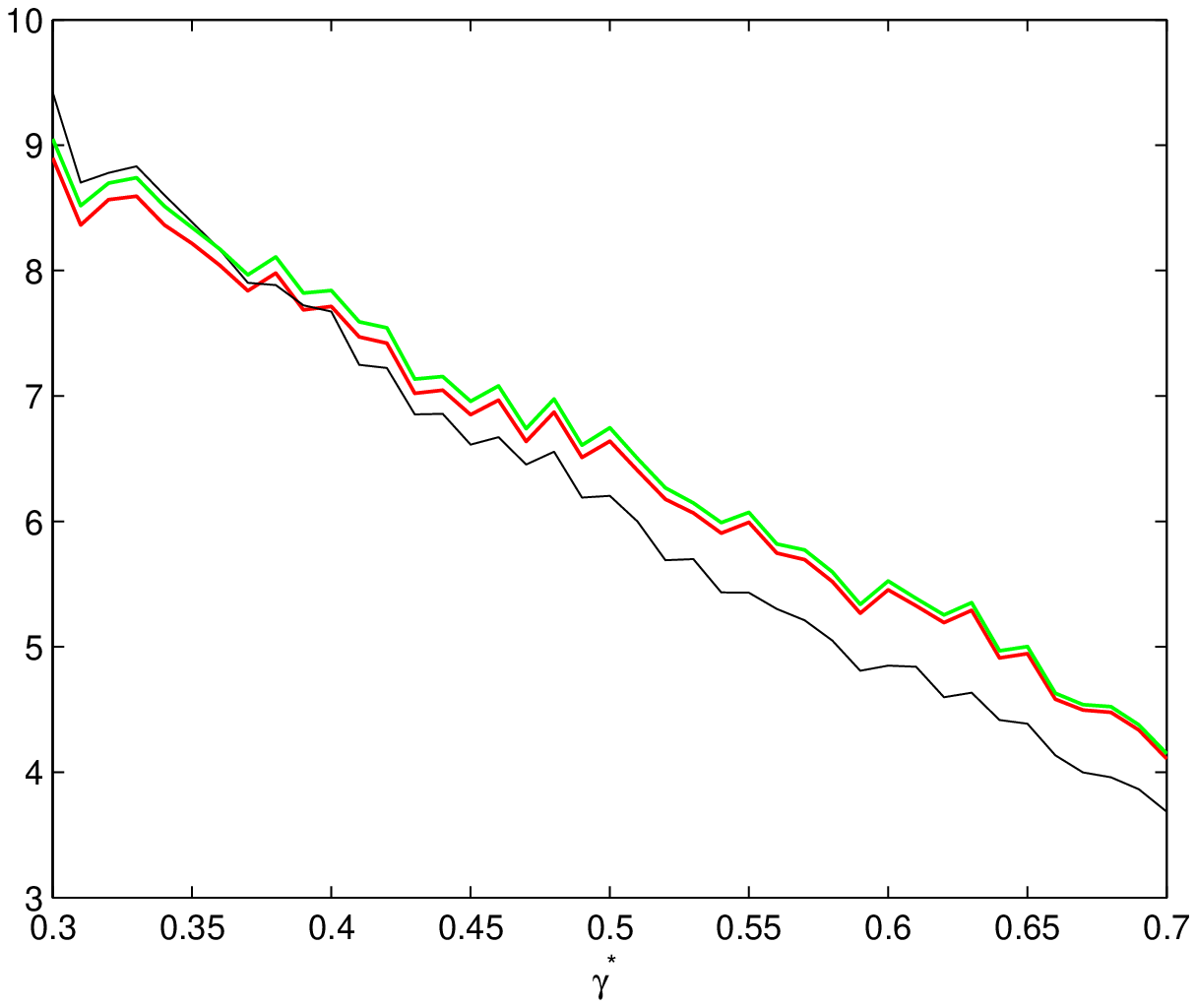}
\includegraphics[height=1.40in]{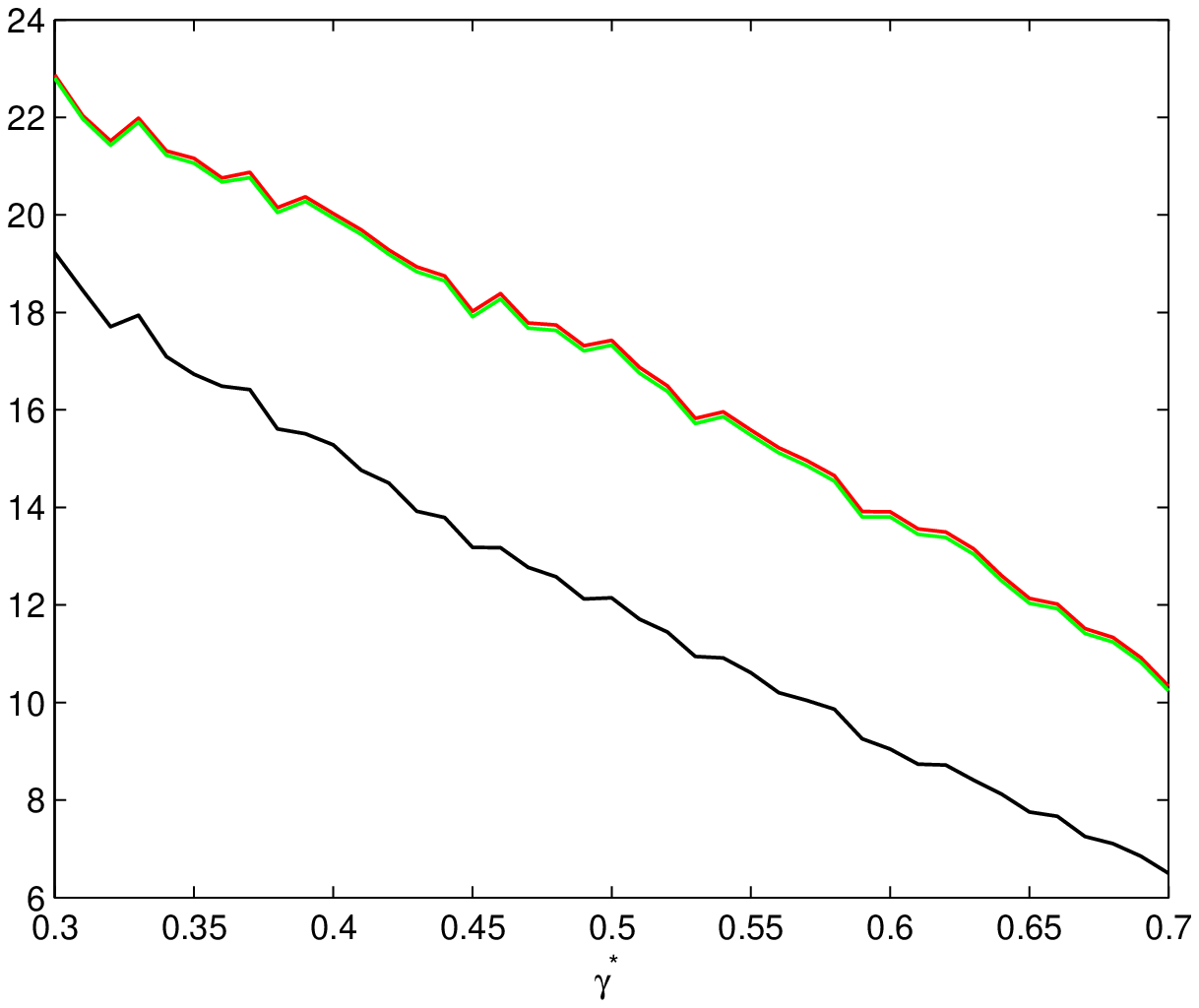}
\includegraphics[height=1.40in]{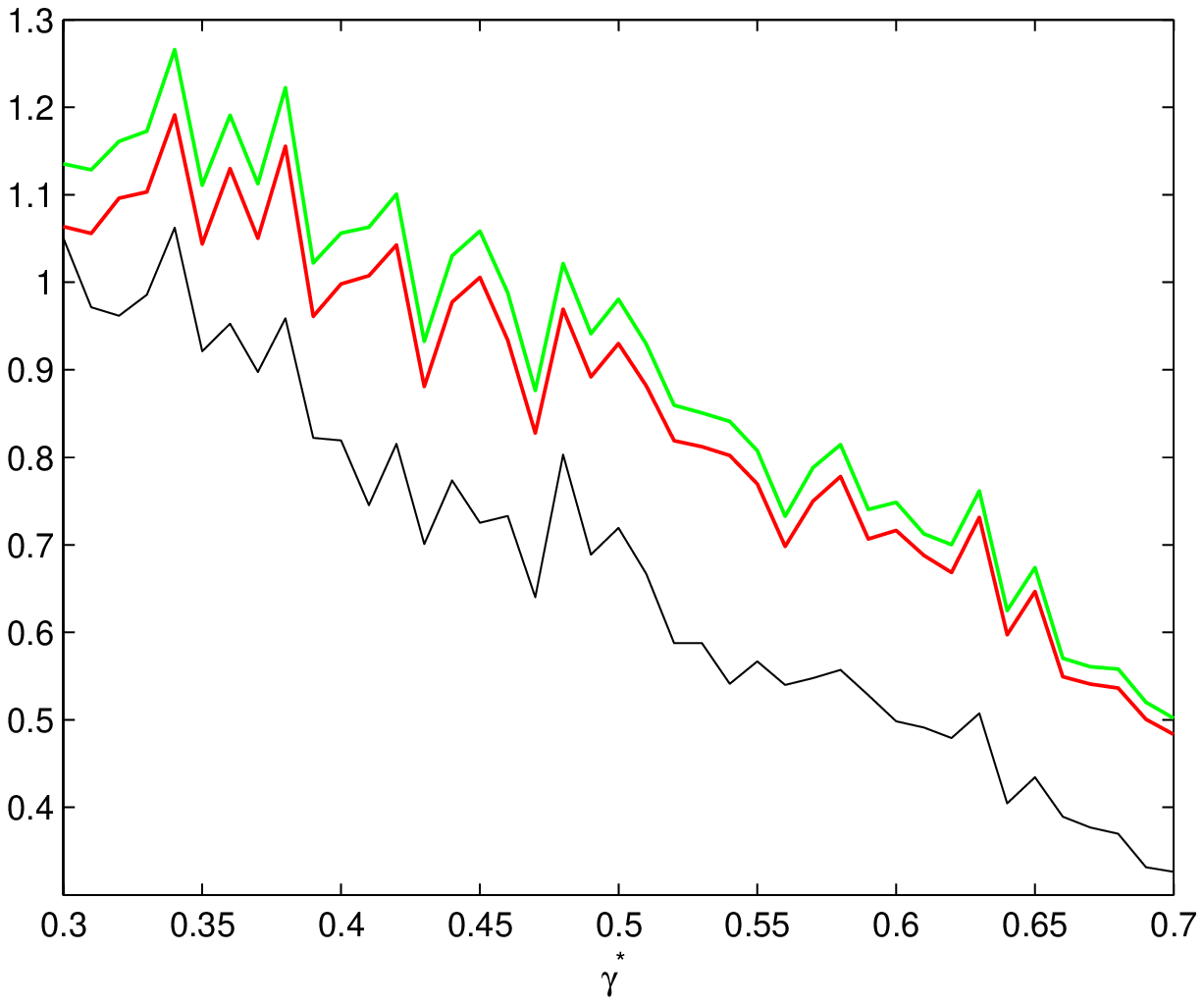}
\includegraphics[height=1.40in]{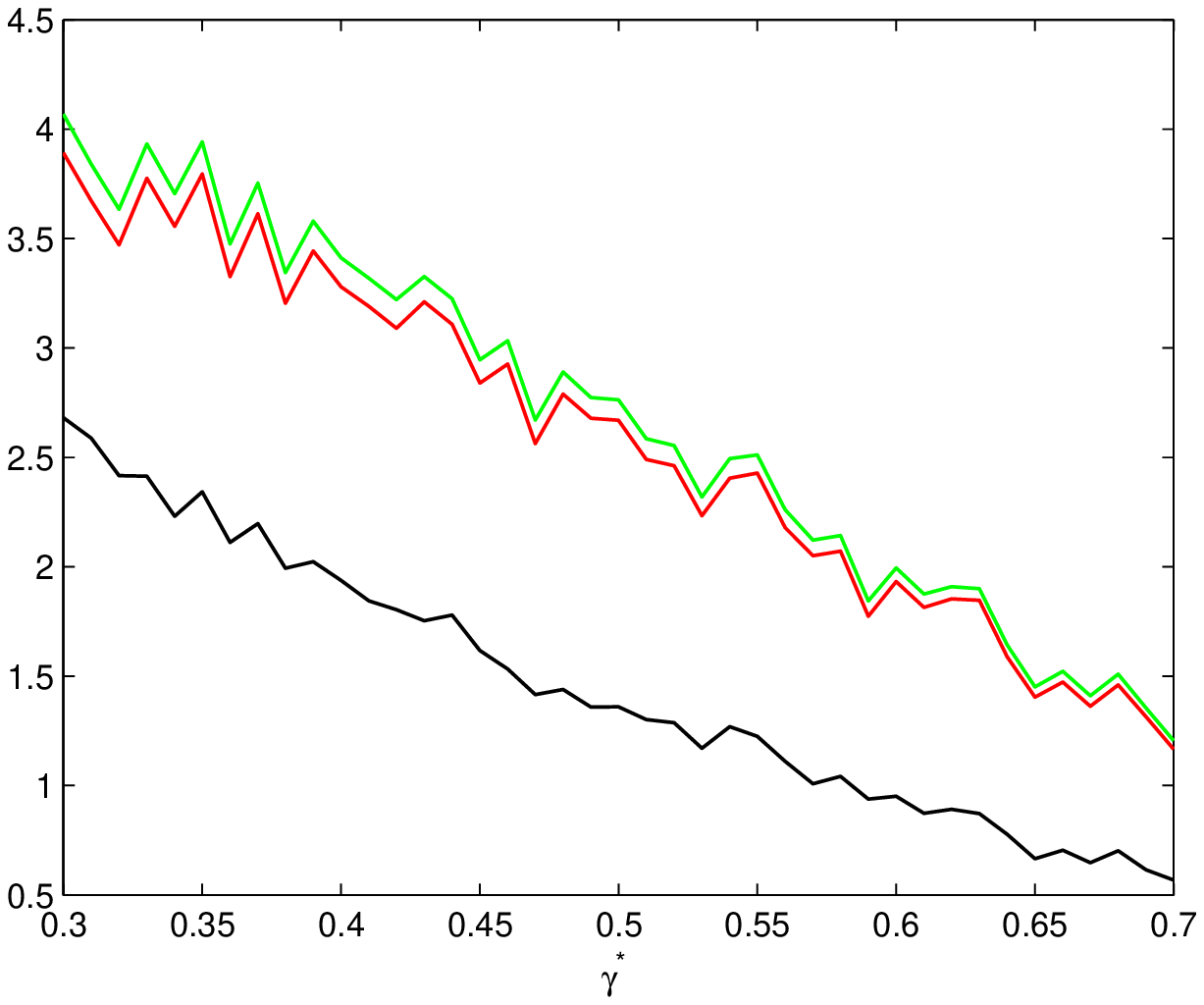}
\end{center}

\textbf{Figure 3:}

\begin{center}
\begin{itemize}
\item
 Comparison of $ E\{L({\boldsymbol{\widehat{\theta}}}_{(\,)}, \boldsymbol{\theta}_{(\,)})\}$
 as a function of $\gamma^*,$ for the predictors\,
$\boldsymbol{\widehat{U}}$,\,\,
$\boldsymbol{\widehat{\theta}}\,^{[2]}_{(\,)}(\sqrt{\gamma^*})$,\,
$\boldsymbol{\widehat{\theta}}\,^{[3]}_{(\,)}$ (red, black, green lines),
where F is the Laplace distribution and G is normal (upper left ) and
where F is the Location exponential distribution and G is normal (upper right)
for, $m=100$.
\item
 Comparison of the MSE of\,
$\widehat{U}_{m}$,\,\,
${\widehat{\theta}_{(m)}}^{\,\,{[2]}}(\sqrt{\gamma^*})$,\,
${\widehat{\theta}_{(m)}}^{\,\,{[3]}}$ (red, black, green lines)
for predicting $\theta_{(m)}$, as a function of $\gamma^*$,
where F is the Laplace distribution and G is normal (bottom left ) and
where F is the Location exponential distribution and G is normal (bottom right),
for $m=100$.
\end{itemize}
\end{center}

In general, the SL estimators and $\boldsymbol{\widehat{\theta}}\,^{[3]}_{(\,)}$ exhibited very similar
performance, see Figure 3. Under the correct assumptions they were somewhat better than our
predictor $\boldsymbol{\widehat{\theta}}\,^{[2]}_{(\,)}(\sqrt{\gamma^*})$ (Figure 2); however, they are usually
computationally intensive and non-robust against model misspecification. Under misspecification of the distributions in the model, it turned out
that $\boldsymbol{\widehat{\theta}}\,^{[2]}_{(\,)}(\sqrt{\gamma^*})$, which does not depend on the assumed model was better, as
can be sees from the   simulations of Figure 3.

\section{Appendix: Proofs}\label{sec:app}
\setcounter{chapter}{7}
\setcounter{equation}{0} 
\noindent {\bf  Proof of Theorem \ref{Th:big}.}
Without loss of generality take $\mu=0$. We have
\begin{align}\label{eq:Ltheta}
&E\{L({\boldsymbol{\widehat{\theta}}}{\,\,^{[2]}_{(\,)}}(\gamma), \boldsymbol{\theta}_{(\,)})\}=
E\sum_{i=1}^m\left(\gamma y_{(i)} +(1-\gamma)
\overline{y}-\theta_{(i)}\right)^2 \nonumber\\
&= E\sum_{i=1}^m\left(\gamma y_{(i)} +(1-\gamma)
\overline{y}-y_{(i)}+y_{(i)}-\theta_{(i)}\right)^2=
E\sum_{i=1}^m\left({y}_{(i)}-\theta_{(i)}\right)^2\\
&+(1-\gamma)^{2}
E\sum_{i=1}^m\left({y}_{(i)}-\overline{y}\right)^2
-2(1-\gamma)E\sum_{i=1}^m\left({y}_{(i)}-\theta_{(i)})({y}_{(i)}-\overline{y}\right)\nonumber.
\end{align}
Therefore,
\begin{align*}
&D(\gamma):=E\{L({\boldsymbol{\widehat{\theta}}{\,\,^{[2]}_{(\,)}}(\gamma),
\boldsymbol{\theta}_{(\,)})}\}- E\{L({\boldsymbol{\widehat{\theta}}{\,\,^{[1]}_{(\,)}},
\boldsymbol{\theta}_{(\,)})}\}=(1-\gamma)^{2}
E\sum_{i=1}^m\left({y}_{(i)}-\overline{y}\right)^2\\
& -2(1-\gamma)
E\sum_{i=1}^m({y}_{(i)}-\theta_{(i)})({y}_{(i)}-\overline{y}).
\end{align*}
 We  calculate  each part  separately. First note that
$E\sum_{i=1}^m\left({y}_{(i)}-\overline{y}\right)^2=
E\sum_{i=1}^m\left({y}_{i}-\overline{y}\right)^2
\\=(\sigma_u^2+\sigma_e^2)(m-1)$.
\ignore{Hence,
\begin{align*}
&D(\gamma)=(\sigma_u^2+\sigma_e^2)(m-1)(1-\gamma)^{2}-2(1-\gamma)
E\sum_{i=1}^m({y}_{(i)}-\theta_{(i)})({y}_{(i)}-\overline{y}).
\end{align*}
}
Now
\begin{align}\label{eq:innprod}
&E\sum_{i=1}^m\left({y}_{(i)}-\theta_{(i)})({y}_{(i)}-\overline{y}\right)=
E\sum_{i=1}^m{y}_{(i)}^{2}-E\sum_{i=1}^m{y}_{(i)}\overline{y}-E\sum_{i=1}^m{\theta_{(i)}}{y}_{(i)}
+E\sum_{i=1}^m{\theta_{(i)}}\overline{y}\nonumber\\
&=m(\sigma_u^2+\sigma_e^2)-m\left(\frac{\sigma_u^2+\sigma_e^2}{m}\right)
-E\sum_{i=1}^m{\theta_{(i)}}{y}_{(i)}
+m\left(\frac{\sigma^{2}_u}{m}\right)=
m(\sigma_u^2+\sigma_e^2)-\sigma_e^2-E\sum_{i=1}^m{\theta_{(i)}}{y}_{(i)}.
\end{align}
Summarizing the above we have
\begin{equation}\label{eq:D}
D(\gamma)=(1-\gamma)^2(\sigma^2_u+\sigma^2_e)(m-1)-
2(1-\gamma)\Big(m(\sigma_u^2+\sigma_e^2)-\sigma_e^2-E\sum_{i=1}^m{\theta_{(i)}}{y}_{(i)}\Big).
\end{equation}
From Lemma \ref{Le:main} (to be proved later)
\begin{align}\label{eq:bounds}
&m(\sigma_u^2+\sigma_e^2)-\sigma_e^2-m\sqrt{\sigma_u^2(\sigma_u^2+\sigma_e^2)}\leq
E\sum_{i=1}^m\left({y}_{(i)}-\theta_{(i)})({y}_{(i)}-\overline{y}\right)
\leq (m-1)\sigma_e^2.
\end{align}
We use the first inequality
to deduce that for $\gamma\leq1$,
\begin{align*}
&D(\gamma)\leq(1-\gamma)^2(\sigma_u^2+\sigma_e^2)(m-1)-2(1-\gamma)\left(m(\sigma_u^2+\sigma_e^2)-
\sigma_e^2-m\sqrt{\sigma_u^2(\sigma_u^2+\sigma_e^2)}\right).
\end{align*}
Equating the right-hand side to zero and solving the quadratic
equation in $1-\gamma$, it is easy to see that $D(\gamma) <0$ in
the interval
$\left(\frac{m}{m-1}(2\sqrt{\gamma^{*}}-1)-\frac{1}{m-1}(2\gamma^{*}-1),
1\right)$, and the result follows. \qed

\noindent {\bf  Proof of Corollary \ref{Cl:c1}}.
Clearly, $E\{L(\boldsymbol{\widehat{\theta}}\,^{[2]}_{(\,)}(\gamma),
\boldsymbol{\theta}_{(\,)})\}\leq E\{L({\boldsymbol{\widehat{\theta}}\,^{[1]}_{(\,)}, \boldsymbol{\theta}_{(\,)})}\}$ for
all $0\leq\gamma\leq1$, if\quad
$\frac{m}{m-1}(2\sqrt{\gamma^{*}}-1)-\frac{1}{m-1}(2\gamma^{*}-1)\leq
0$. Solving the quadratic equation in $\sqrt{\gamma^{*}}$, we see that
the latter inequality holds if either $(i)\,\,\,
\gamma^{*}\leq\left(\frac{m-\sqrt{(m-1)^2+1}}{2}\right)^2$ or
$(ii)\,\,\,
\gamma^{*}\geq\left(\frac{m+\sqrt{(m-1)^2+1}}{2}\right)^2 (\geq 1)$.
Since $\gamma^{*}\leq 1$ the only possibility is $(i)$,
and the proof is complete.\qed

\noindent {\bf  Proof of Corollary \ref{Cl:c2}}.
From Theorem \ref{Le:main} it is clear that
$E\{L(\boldsymbol{\widehat{\theta}}\,^{[2]}_{(\,)}(\gamma),
\boldsymbol{\theta}_{(\,)})\}\leq E\{L({\boldsymbol{\widehat{\theta}}\,^{[1]}_{(\,)}, \boldsymbol{\theta}_{(\,)})}\}$ if
$\frac{m}{m-1}(2\sqrt{\gamma^{*}}-1)-\frac{1}{m-1}(2\gamma^{*}-1)\leq\gamma^{*}
\leq \gamma \leq 1$. The first inequality is equivalent to
$\gamma^{*}\leq\frac{(m-1)^2}{(m+1)^2}$ or $\gamma^{*}\geq 1$. The
case $\gamma^{*}=1$ is trivial because in this case
${\widehat{\theta}_{(i)}}^{\,\,[1]}=
{\widehat{\theta}_{(i)}}^{\,\,[2]}(\gamma^*)$. \qed

\noindent {\bf   Proof of Lemma  \ref{Le:main}}.
The lower bound is a result of the rearrangement inequality
\begin{align*}
&E\sum_{i=1}^m{\theta_{(i)}}{y}_{(i)}\geq
E\sum_{i=1}^m{\theta_{i}}{y}_{i}=m(\sigma_u^2+\mu^2).
\end{align*}
The upper bound follows from
\begin{align*}
E\sum_{i=1}^m{\theta_{(i)}}{y}_{(i)}\leq
\left(
\sum_{i=1}^mE({\theta_{i}^{2}})\sum_{i=1}^m{E({y}_{i}^{2}})\right)^{1/2}
=m\sqrt{(\sigma_u^2+\mu^2)(\sigma_u^2+\sigma_e^2+\mu^2)},
\end{align*}
where the inequality follows from the Cauchy-Schwarz
inequality. \qed

\noindent {\bf  Proof of Theorem \ref{Th:MinPoint}}. By the
calculations of  Theorem \ref{Th:big},
\begin{align*}
&E\{L(\boldsymbol{\widehat{\theta}}\,^{[2]}_{(\,)}(\gamma),
\boldsymbol{\theta}_{(\,)})\}=
E\sum_{i=1}^m\left({y}_{(i)}-\theta_{(i)}\right)^2\\
&+(1-\gamma)^{2} (\sigma_u^2+\sigma_e^2)(m-1)
-2(1-\gamma)E\sum_{i=1}^m\left({y}_{(i)}-\theta_{(i)})({y}_{(i)}-\overline{y}\right).
\end{align*}
Hence, $d E\{L(\boldsymbol{\widehat{\theta}}\,^{[2]}_{(\,)}(\gamma),
\boldsymbol{\theta}_{(\,)})\}/d{\gamma}=0$ if and only if
$\gamma=1-\frac{E\sum_{i=1}^m\left({y}_{(i)}
-\theta_{(i)})({y}_{(i)}-\overline{y}\right)}{(m-1)(\sigma_u^2+\sigma_e^2)}$,
which is a minimum by convexity. We cannot calculate the latter
expression exactly, yet the bounds of   (\ref{eq:bounds}) imply the
result readily. \qed

\noindent {\large\bf Acknowledgment} We thank Danny Pfeffermann for
discussions that led to the formulation of the problems studied in
this paper. An Associate Editor and the referees made comments that resulted in significant
improvements in the paper.\\ This paper is dedicated to the memory of Gideon Schwarz, a teacher and a friend.

This research was supported in part by grant number 473/04 from the Israel Science
Foundation.
\\
\\
\noindent{\Large\bf References}
\begin{description}
\item Battese, G. E., Harter, R. M., and Fuller, W. A. (1988). An error component model
for predection of county crop areas using survey and satellite data.
{\it Journal of the American Statistical
Association} {\bf 83}, 28-36.

\item Blumenthal, S. and Cohen, A.  (1968). Estimation of the
larger of two normal means. {\it Journal of the American Statistical
Association} {\bf 63}, 861-876.

\item David, H. A. and Nagaraja, N. H. (2003). \textit{Order Statistics}
(third edition). {Wiley, New York.}

\item Dawid, A. P. (1994). Selection paradoxes of Bayesian inference.
{\it In Multivariate Analysis and its Application} {\bf 24},
(eds. T.W. Anderson, K. A-T. A Fang and I. Olkin)
{Philadelphia, PA:IMS}.

 \item Fay, R. E.
and Herriot, R. A.  (1979). Estimates of income for small places: An
application of James-Stein procedures to census data. {\it Journal
of the American Statistical Association} {\bf 74}, 269-277 .

\item Ghosh, M. (1992). Constrained Bayes estimates with application. {\it Journal
of the American Statistical Association} {\bf 87}, 533-540.

\ignore{ \item Henderson, C. R. (1975). Best Linear Unbiased Estimation and
 Prediction Under a Selection Model. {\it Biometrics} {\bf 31},
423-447.}

\item Kella, O. (1986). On the distribution of the maximum
of bivariate normal random variables with general means and
variances. {\it Commun. Statist-Theory Meth.} {\bf 15}, 3265-76.

\item Louis, T. A. (1984). Estimating a population of parameter values using Bayes and empirical
Bayes methods. {\it Journal
of the American Statistical Association} {\bf 79}, 393-398.


\item
Pfeffermann, D. (2002). Small area estimation- new developments and directions. {\it International Statistical Review} {\bf 70}, 125-143.

\item
  Rao, J. N. K.  (2003). \textit{Small Area Estimation}. Wiley, New York.

\item Robinson, G. K. (1991). That BLUP is a good thing: the estimation of random effects.
{\it Statistical Science} {\bf 6}, 15-32.

 \item
  Rinott, Y. and
Samuel-Cahn, E.  (1994). Covariance between variables and their
order statistics for multivariate normal variables. {\it Statist.
Probab. Lett. } {\bf 21}, 153-155.

\item
Senn, S. (2008). A Note concerning a selection "Paradox" of Dawid's.
{\it The American Statistician } {\bf 62}, 206-210.
\item
Shen, W. and Louis, T. A. (1998). Triple-goal estimates in two-stage hierarchical models.
{\it Journal of the Royal Statistical Society B } {\bf 60}, 455-471.
\item
Shen, W. and Louis, T. A. (2000). Triple-Goal estimates for Disease Mapping.
{\it Statistics in Medicine} {\bf 19}, 2295-2308.

\item Schwarz, G. (1987). A minimax property of linear regression.
{\it Journal of the American Statistical Association} {\bf 82},
220.

\item Siegel, A. F. (1993). A surprising covariance involving the
minimum of multivariate normal variables. {\it Journal of the
American Statistical Association} {\bf 88}, 77-80.

\item Stein, C. (1956). Inadmissibility of the usual estimator for the mean of a
multivariate normal distribution. {\it Proc. Third Berkeley Symp. Math. Statist. Probability
{\bf  1} }, 197-206. University of California Press, Berkeley, CA.

\item Wright, D. L., Stern, H. S., and  Cressie, N. (2003). Loss function for estimation of extreme
with an application to disease mapping. {\it The Canadian Journal of Statistics } {\bf 31}, 251-266.

\end{description}


\noindent first author affiliation \vskip 2pt
\noindent E-mail: (msyakov@mscc.huji.ac.il) \vskip 1pt
\noindent second author affiliation \vskip 1pt \noindent E-mail:
(rinott@mscc.huji.ac.il)

\par
\newpage
\centerline{\large\bf A Supplement to Prediction of Ordered Random Effects in a
Simple Small Area Model}
\vspace{.1cm}
In this Supplement we provide some of the simulations and technical proofs.
Equations in this Supplement are indicated by $S$, e.g., $(3.1S)$, and similarly, lemmas
that appear only in the Supplement are numbered with $S$, e.g., Lemma $1S$. Equations, lemmas, and Theorems
without  $S$, refer to the article itself. Most of the notation is defined in the article, and this Supplement
cannot be read independently.
\section{Simulations for Conjecture \ref{Conj:C1}}
\begin{Conjecture}
\label{Conj:C1}
 The optimal $\gamma$ in the sense of Theorem
4, $\gamma^{\,o}$, satisfies
$$\lim_{m\rightarrow\infty}\gamma^{\,o}=\sqrt{\gamma^{*}}.$$
\end{Conjecture}
We justify  Conjecture \ref{Conj:C1} by simulations. First we consider the case that
both the area random effect $u_i$ and the sampling error $e_i$ have a
normal distributions, and take  $m=5, 10, 20, 100$, and then repeat the simulation
with $e_i$ having a translated exponential distribution.
The red lines in Figure 1S are the range (3.6) of optimal $\gamma$
from Theorem 4 and the blue line is the optimal
$\gamma$, both as functions of $\gamma^*$. The
simulations were done as follows: we set $\sigma^2_{u}=1$. Different
values of $\sigma^2_{e}$ define the different values of $\gamma^*$.
Setting without loss of generality $\mu=0$, we generated $y_i=0+u_i+e_i$, $i=1,\ldots,m$. For
each value of $\gamma^*$ we ran 1,000 simulations. By suitably averaging over these simulations, we then
approximated $E\{L(\boldsymbol{\widehat{\theta}}\,^{[2]}_{(\,)}(\gamma),
\boldsymbol{\theta}_{(\,)})\}$
for each $\gamma\in[0,1]$ using an exhaustive search with step-size
of 0.001 and found $\gamma^{\,o}$, the value of $\gamma$ that minimizes
 $E\{L(\boldsymbol{\widehat{\theta}}\,^{[2]}_{(\,)}(\gamma),
\boldsymbol{\theta}_{(\,)})\}$.
\begin{center}
\includegraphics[height=2.1in, width=2.6in]{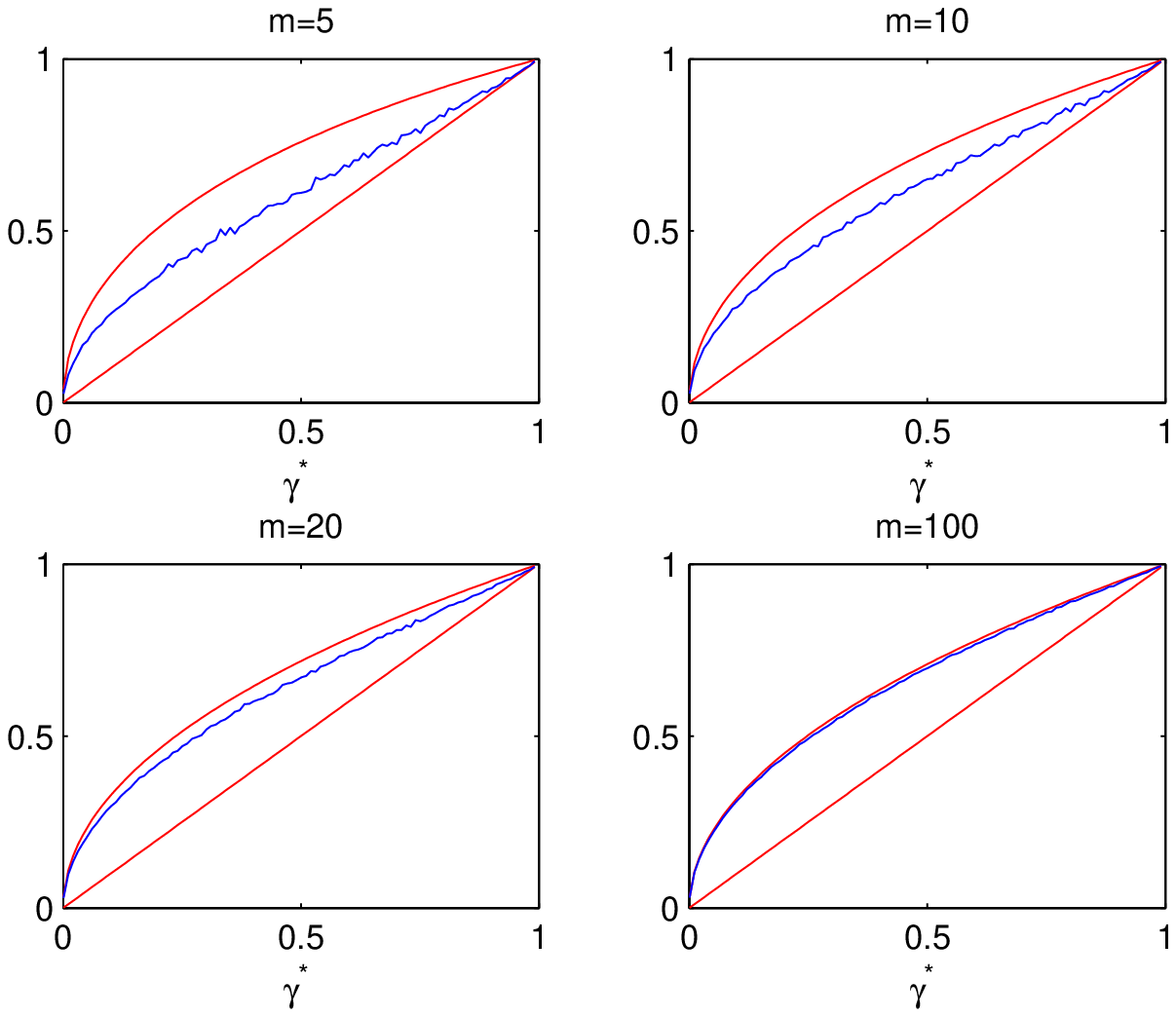}
\includegraphics[height=2.1in, width=2.6in]{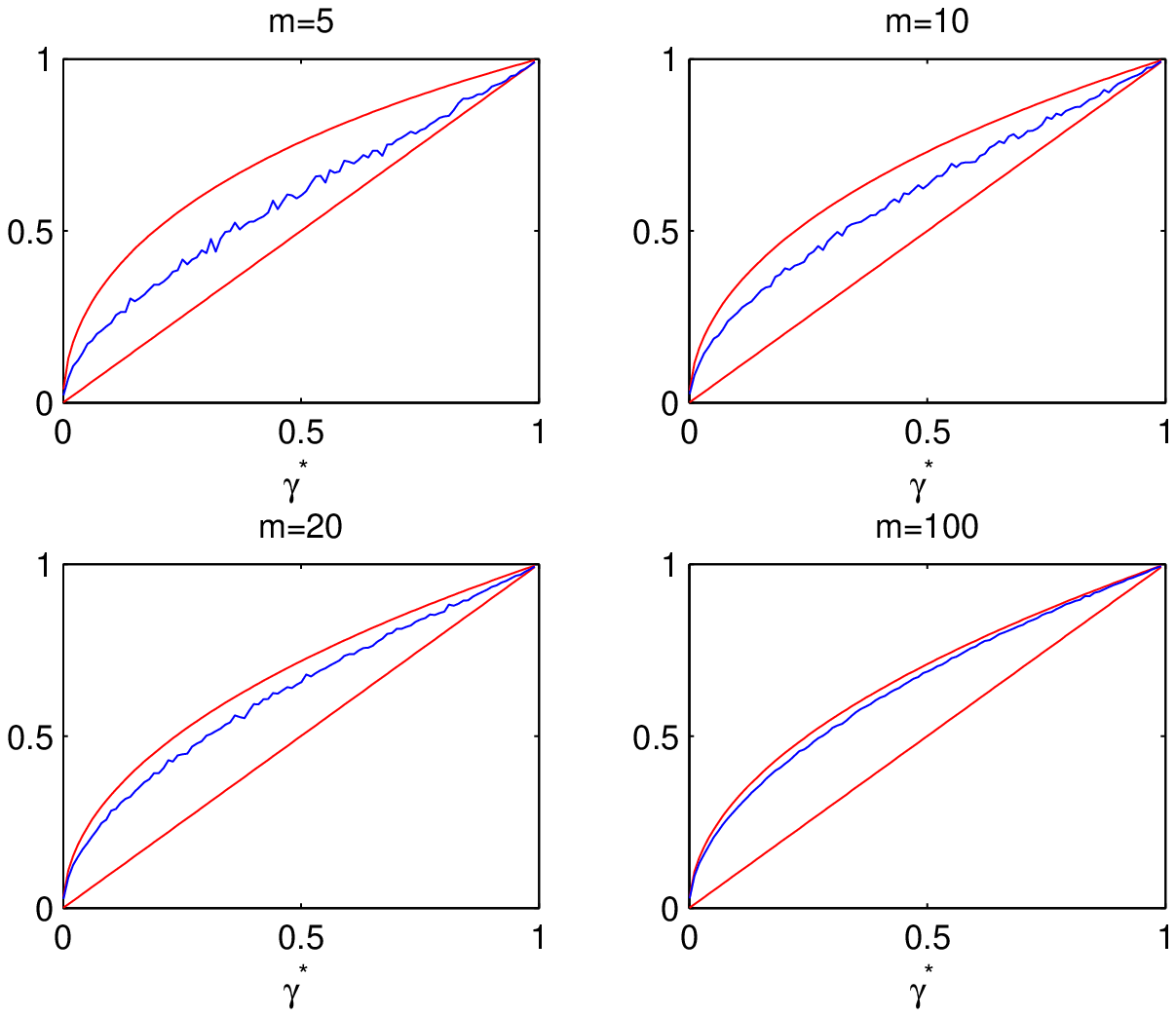}

\textbf{Figure 1S:} $\gamma^{\,o}$ (the optimal $\gamma$) as a function of $\gamma^*$
(blue line) and the range of  optimal $\gamma$ from Theorem
4  as a function of $\gamma^*$ (red lines) when:
\begin{enumerate}
 \item Both the area random effect $u_i$ and the sampling error
$e_i$ are normal (left four graphs).
\item The area
random effects $u_i$ are normal, but the sampling errors $e_i$ are
from a location exponential distribution (an exponential
distribution translated by a constant) (right four graphs).
\end{enumerate}
\end{center}

\section{Simulations for Conjecture 2, and comparison of predictors}\label{sec:sim}
\subsection{ Known variances }
For normal F and G, Conjecture 2
says that the predictor ${\widehat{\theta}_{(i)}}^{\,\,{[3]}}$ is
better than ${\widehat{\theta}_{(i)}}^{\,\,{[2]}}(\gamma)$ for all
values of $\gamma$ (including the optimal) in the sense that
$E\{L(\boldsymbol{\widehat{\theta}}\,^{[3]}_{(\,)},
\boldsymbol{\theta}_{(\,)})\} \le
E\{L(\boldsymbol{\widehat{\theta}}\,^{[2]}_{(\,)}(\gamma),
\boldsymbol{\theta}_{(\,)})\}$. Recall that
$E\{L(\boldsymbol{\widehat{\theta}}\,^{[2]}_{(\,)}(\gamma^o),
\boldsymbol{\theta}_{(\,)})\} \le
E\{L(\boldsymbol{\widehat{\theta}}\,^{[2]}_{(\,)}(\gamma),
\boldsymbol{\theta}_{(\,)})\}$  for all
$\gamma$.

The simulations below support Conjecture 2.    Figure 2S shows a sample of simulation
results for  $m=30$ and 100.  We compare the expected loss in predicting
$\boldsymbol{\theta}_{(\,)}$ by
$\boldsymbol{\widehat{\theta}}\,^{[2]}_{(\,)}(\gamma^{\,o})$ to that of
$\boldsymbol{\widehat{\theta}}\,^{[3]}_{(\,)}$. While doing these simulations, we also compared the expected
loss  in predicting
 $\theta_{(m)}$ by
${\widehat{\theta}_{(m)}}^{\,\,{[2]}}(\gamma^{\,o})$ to that of
${\widehat{\theta}_{(m)}}^{\,\,{[3]}}$.

The simulations  show that  the expected losses of the predictors
$\boldsymbol{\widehat{\theta}}\,^{[2]}_{(\,)}(\gamma^{\,o})$ and
$\boldsymbol{\widehat{\theta}}\,^{[3]}_{(\,)}$ are rather close, while the predictor
$\boldsymbol{\widehat{\theta}}\,^{[2]}_{(\,)}(\gamma^{*})$  is far
worse. This suggests    that the linear predictor
$\boldsymbol{\widehat{\theta}}\,^{[2]}_{(\,)}(\gamma^{\,o})$  can be
used without much loss.  It is important to note that given $\gamma^{\,o}$ this
estimator is easy to calculate. For large $m$ one may take $\gamma^{\,o}=\sqrt{\gamma^{*}}$, whereas
for small $m$, the approximation of Section 3.2  can be used.


The simulation was done as
follows: we set $\sigma^2_{u}=1$. Different values of $\sigma^2_{e}$
define the different values of $\gamma^*$. Setting $\mu=0$, we
generated $y_i=0+u_i+e_i$, $i=1,..,m$. For each value of $\gamma^*$
we ran 1,000 simulations and approximated
$E\{L({\boldsymbol{\widehat{\theta}}\,^{[2]}_{(\,)}(\gamma), \boldsymbol{{\theta}}_{(\,)})}\}$ for each
$\gamma$ in the range (3.6). Using an exhaustive search
with step-size of 0.001 we found $\gamma^{\,o}$, the minimizer of
$E\{L({\boldsymbol{\widehat{\theta}}\,^{[2]}_{(\,)}(\gamma), \boldsymbol{{\theta}}_{(\,)})}\}$.
We approximated ${\widehat{\theta}_{(i)}}^{\,\,{[3]}}$ in
the following way: when both F and G are normal,
$\theta_{i}|y_i~\thicksim
N\left(\gamma^*y_i+(1-\gamma^*)\mu, \gamma^*\sigma^2_{e}\right).$
Hence, for each $y_{i},\, i=1,...,m,$ we generated 1,000 random variables
from $N\left(\gamma^*y_i+(1-\gamma^*)\overline{y},
\gamma^*\sigma^2_{e}\right)$, sorted them, and approximated ${\widehat{\theta}_{(i)}}^{\,\,{[3]}}.$
We approximated $E\{L({\boldsymbol{\widehat{\theta}}\,^{[3]}_{(\,)}, \boldsymbol{{\theta}}_{(\,)})}\}$ in
the same way as we approximated
$E\{L({\boldsymbol{\widehat{\theta}}\,^{[2]}_{(\,)}(\gamma^*), \boldsymbol{{\theta}}_{(\,)})}\}.$\\

\begin{center}
\includegraphics[height=1.50in]{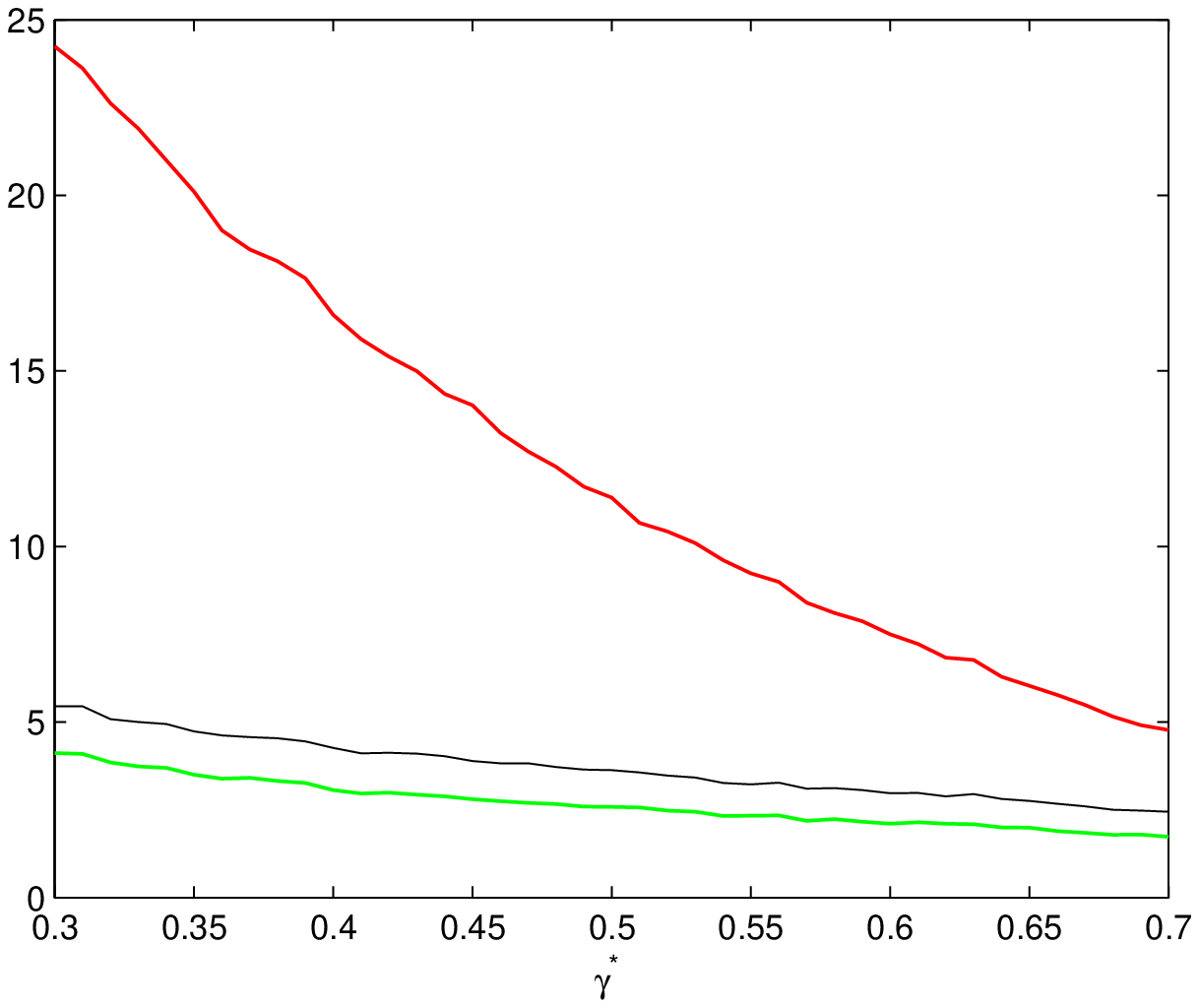}
\includegraphics[height=1.50in]{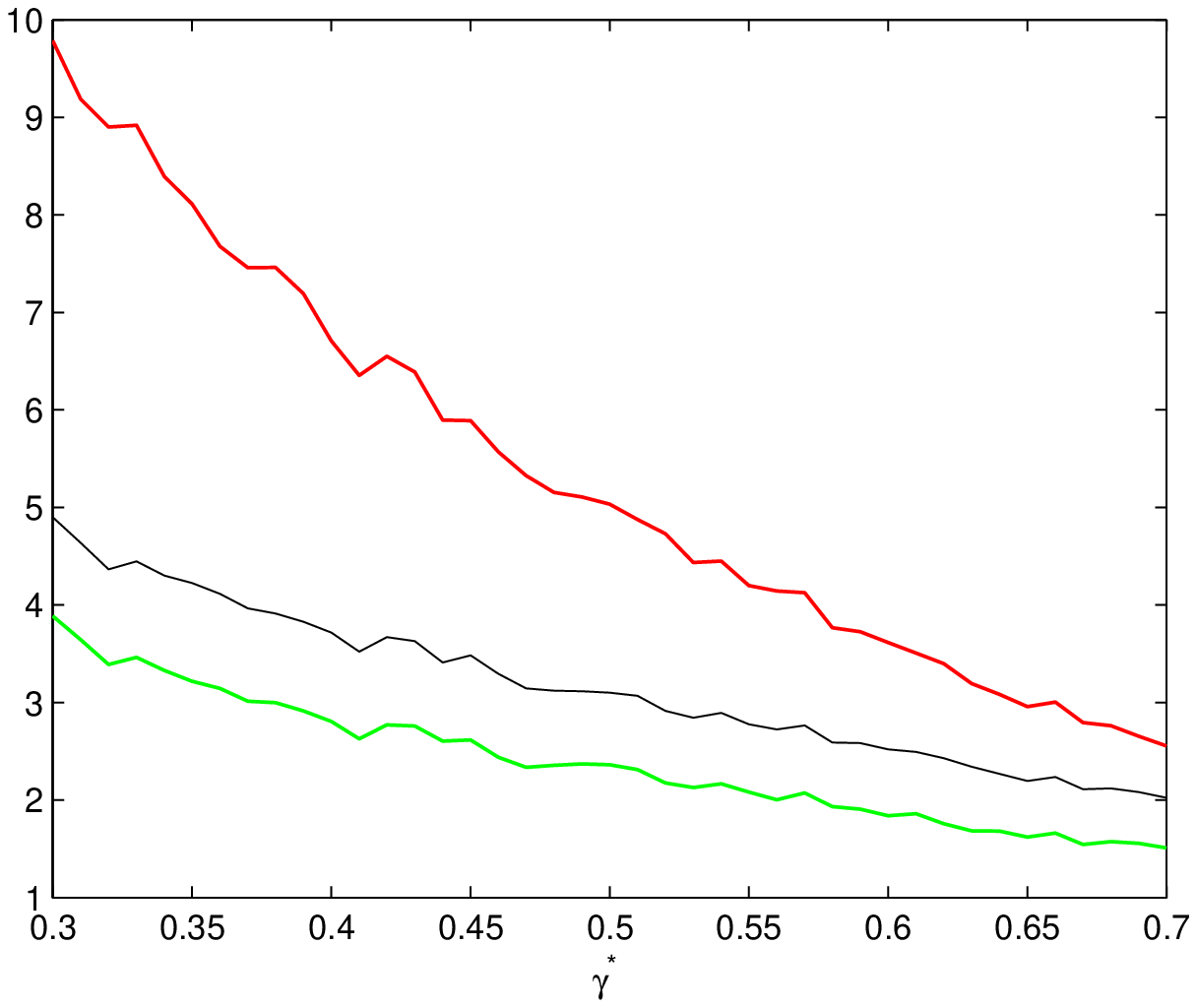}
\includegraphics[height=1.50in]{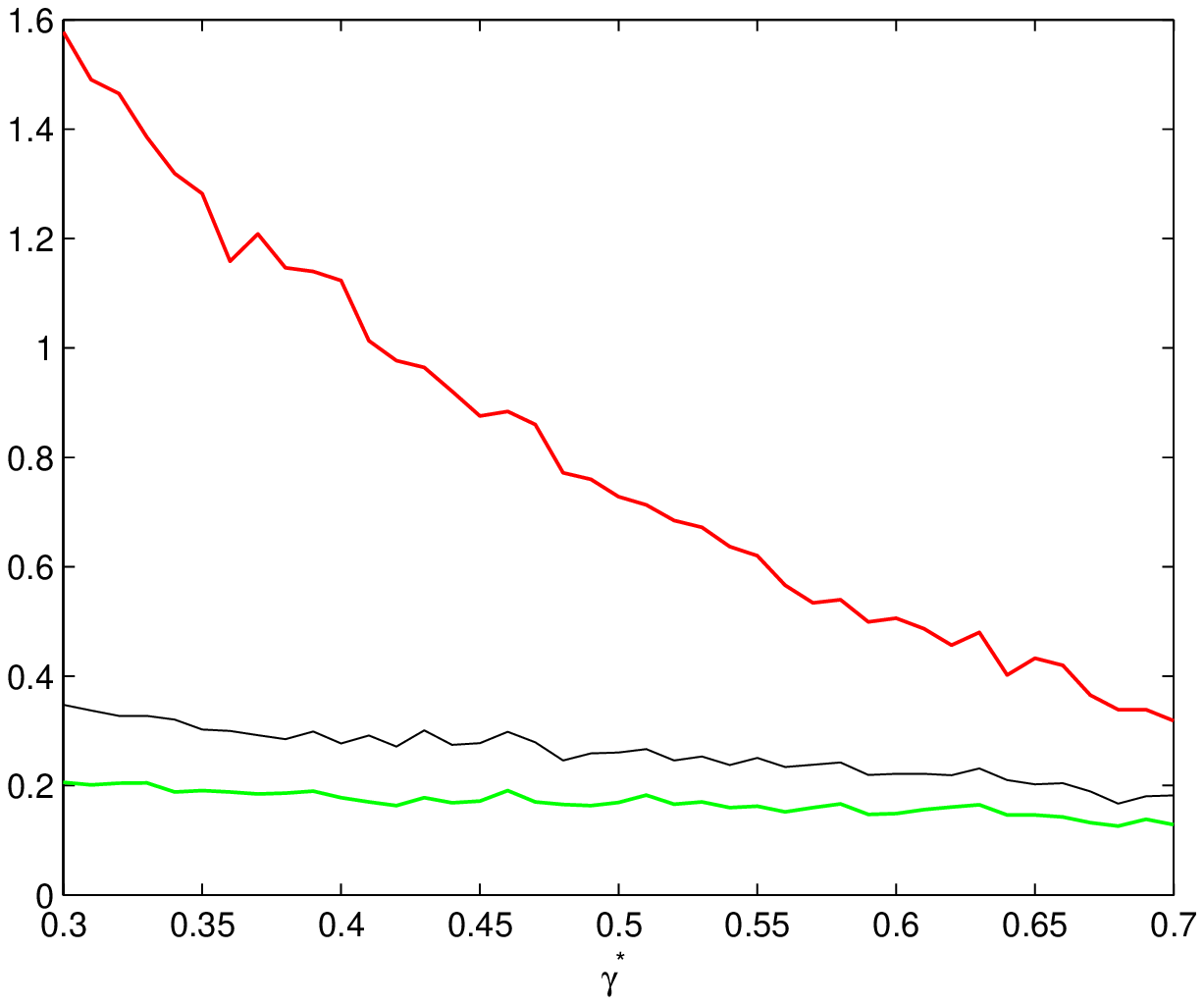}
\includegraphics[height=1.50in]{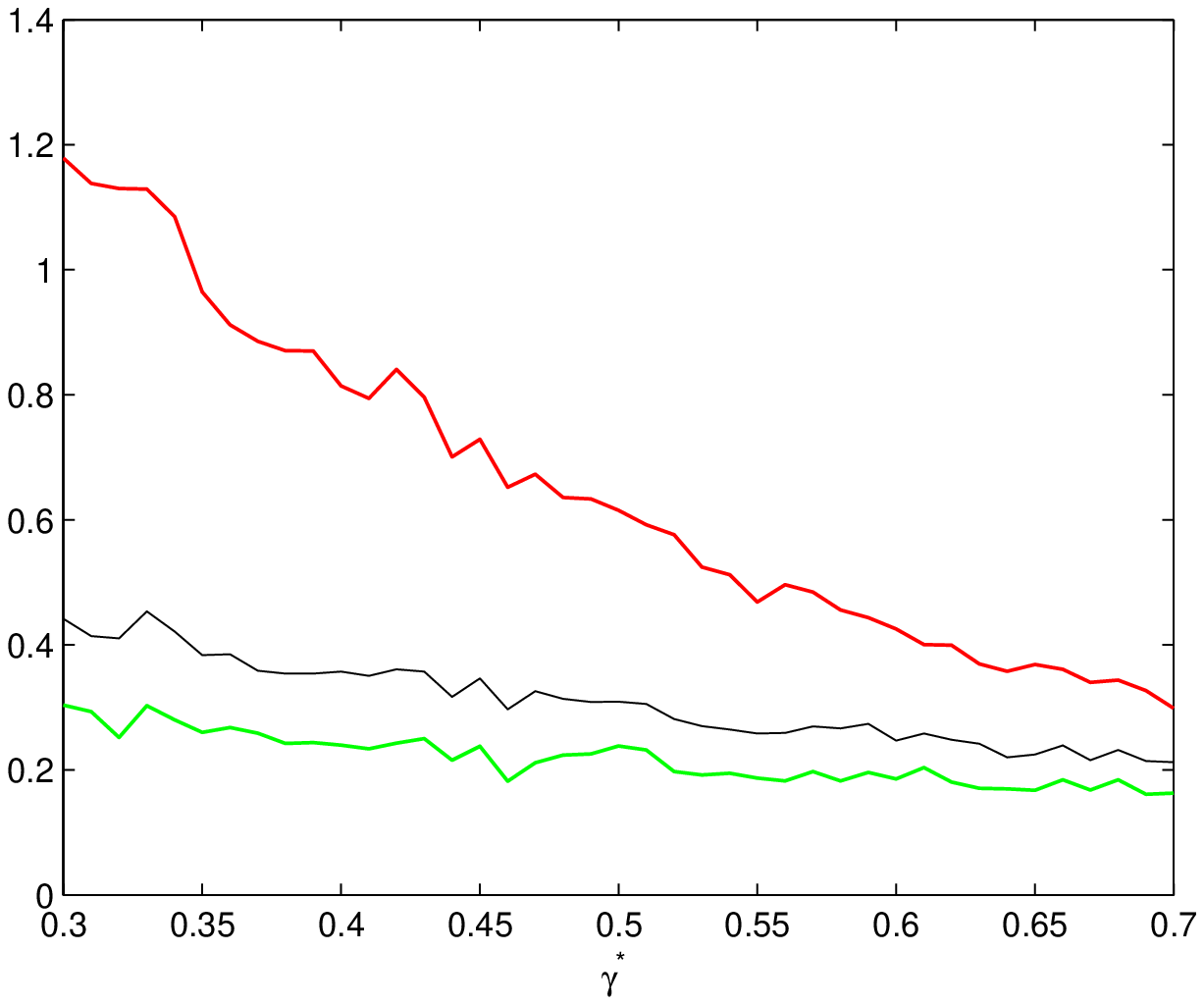}
\end{center}

\textbf{Figure 2S:}
\begin{center}
\begin{itemize}
\item
 Comparison of $ E\{L(\boldsymbol{\widehat{\theta}}_{(\,)}, \boldsymbol{{\theta}}_{(\,)})\}$
 as a function of $\gamma^*,$ for the predictors
$\boldsymbol{\widehat{\theta}}\,^{[2]}_{(\,)}(\gamma^{\,*})$,\,
$\boldsymbol{\widehat{\theta}}\,^{[2]}_{(\,)}(\gamma^{\,o})$,\,
$\boldsymbol{\widehat{\theta}}\,^{[3]}_{(\,)}$ (red, black, green lines),
where F and G are normal and\\ $m=100$ (upper left ), $m=30$ (upper right)
\item
 Comparison of the MSE of
${\widehat{\theta}_{(m)}}^{\,\,{[2]}}(\gamma^*),\,
{\widehat{\theta}_{(m)}}^{\,\,{[2]}}(\gamma^{\,o}),\,
{\widehat{\theta}_{(m)}}^{\,\,{[3]}}$ (red, black, green lines) for predicting
$\theta_{(m)}$, as a function of $\gamma^*$, where F and G are
normal and\\ $m=100$ (bottom left), $m=30$ (bottom right)
\end{itemize}
\end{center}

\subsection{Unknown variances} Figure $3S$ compares the risks when only
$\sigma_u^2$ is unknown and its estimator  (4.1)  is plugged-in. Otherwise,
the simulations are similar to those of the previous section.
The case that both variances, $\sigma_u^2$
and $\sigma_e^2$ are unknown is considered in Section 4 in the article.
\begin{center}
\includegraphics[height=1.50in]
{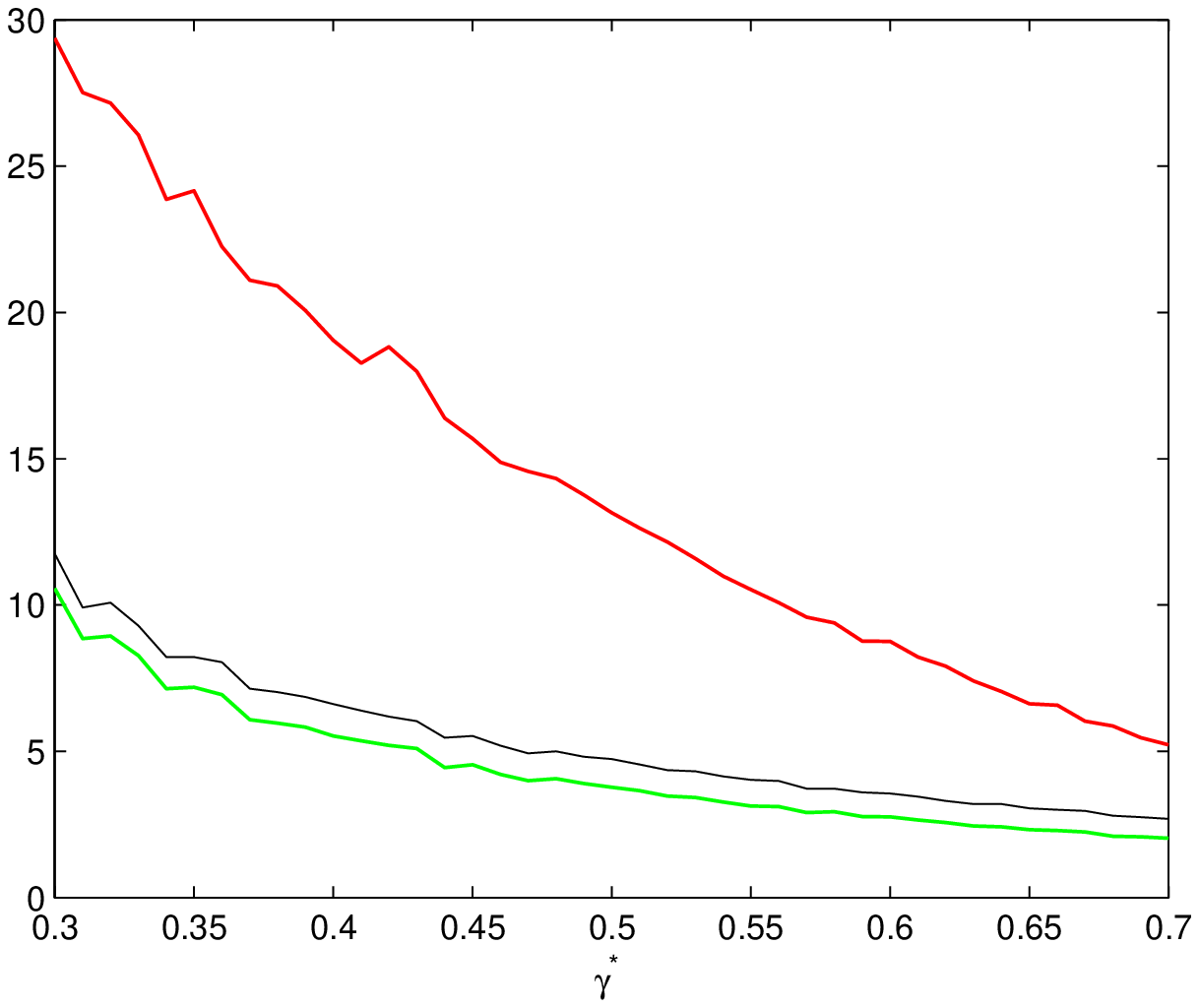}
\includegraphics[height=1.50in]
{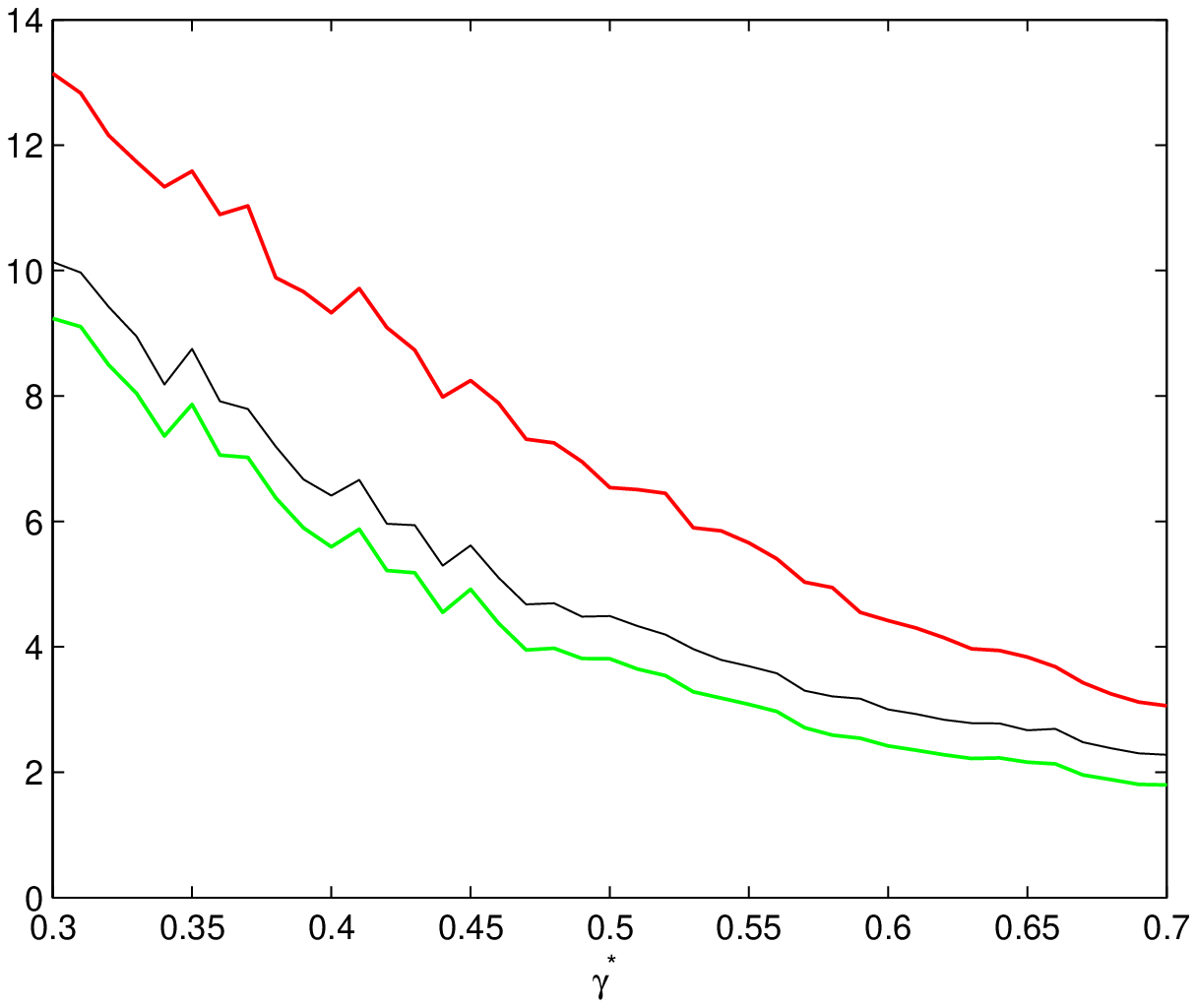}

\includegraphics[height=1.50in]{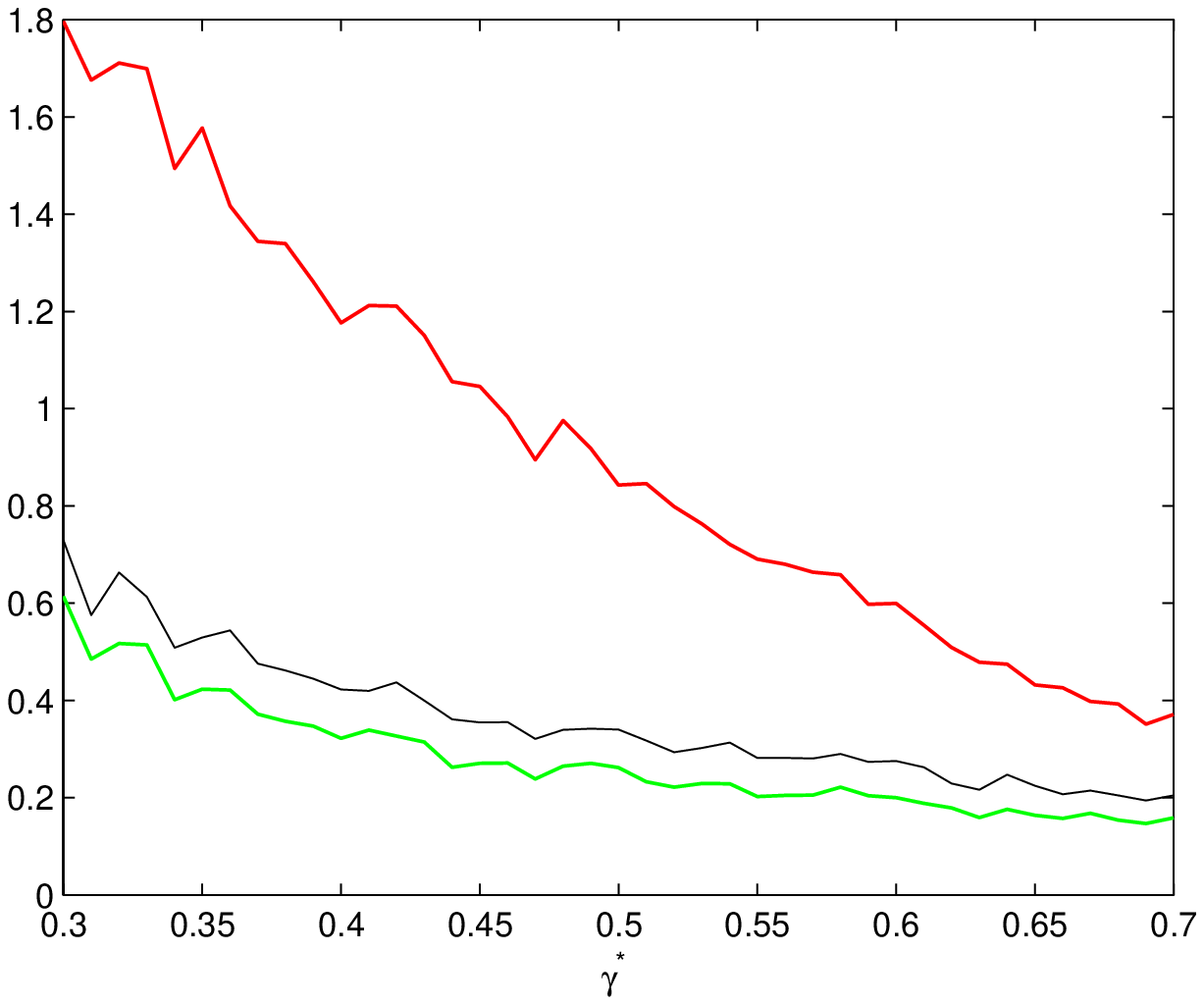}
\includegraphics[height=1.50in]{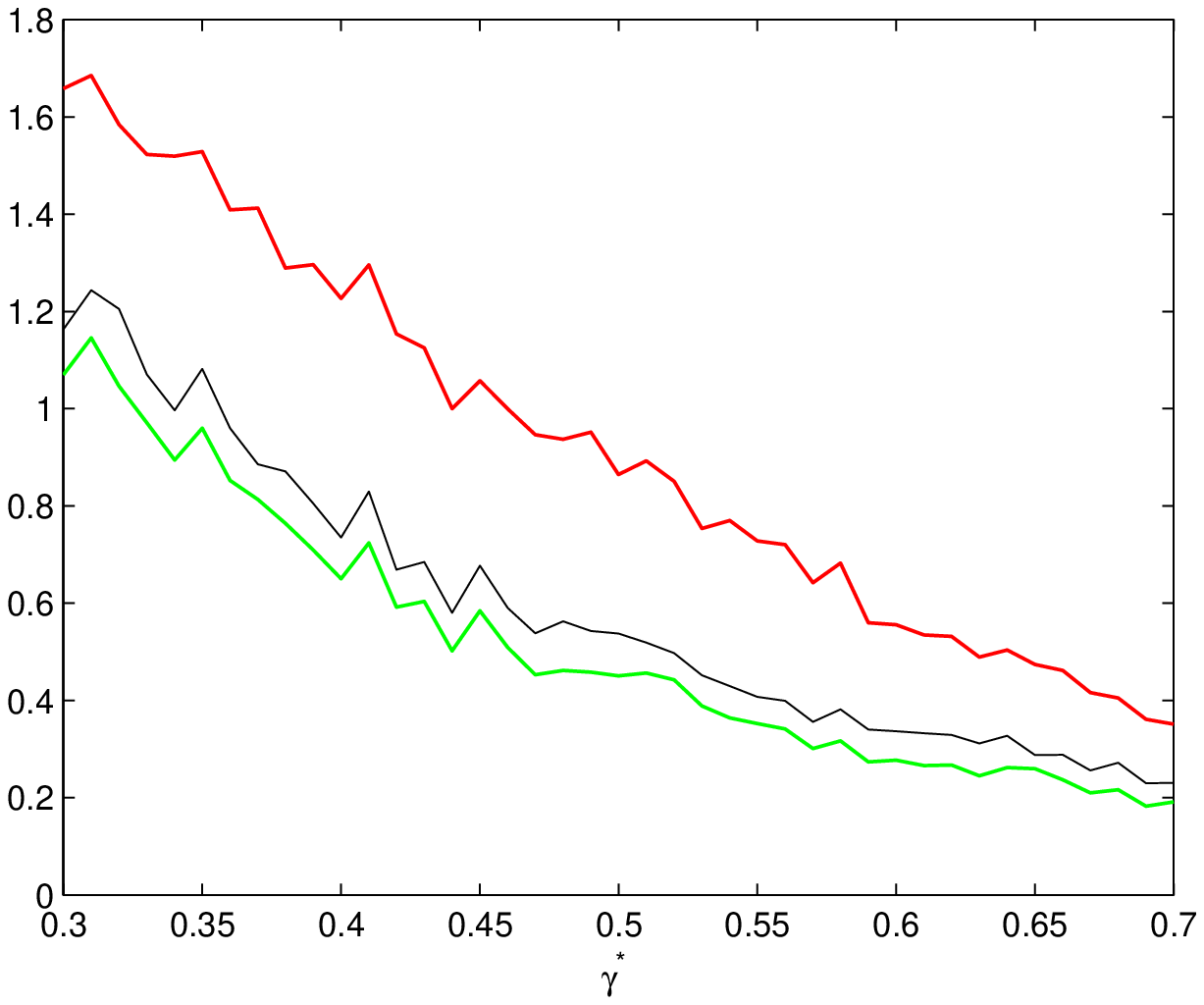}
\end{center}

\textbf{Figure 3S:}
\begin{center}
\begin{itemize}
\item
 Comparison of $ E\{L(\boldsymbol{\widehat{\theta}}_{(\,)}, \boldsymbol{{\theta}}_{(\,)})\}$
 as a function of $\gamma^*,$ for the predictors
$\boldsymbol{\widehat{\theta}}\,^{[2]}_{(\,)}(\gamma^{\,*})$,\,
$\boldsymbol{\widehat{\theta}}\,^{[2]}_{(\,)}(\sqrt{\gamma^{\,*}})$,\,
$\boldsymbol{\widehat{\theta}}\,^{[3]}_{(\,)}$ (red, black, green lines),
where F and G are normal and\\ $m=100$ (upper left ), $m=30$ (upper right)
\item
Comparison of the MSE of
${\widehat{\theta}_{(m)}}^{\,\,{[2]}}(\gamma^*),\,
{\widehat{\theta}_{(m)}}^{\,\,{[2]}}(\sqrt{\gamma^*}),\,
{\widehat{\theta}_{(m)}}^{\,\,{[3]}}$ (red, black, green lines)
for predicting $\theta_{(m)}$, as a function of $\gamma^*$, where F and G
are normal and  $m=100$ (bottom left), $m=30$ (bottom right)
\end{itemize}
\end{center}

\section{Proofs}\label{sec:app}
\setcounter{chapter}{3}
\subsection{Proof of Theorem 5}


For the proof of Theorem 5 
we need some further lemmas.
In the sequel, $\mathbb{I}$ denotes an indicator function, and
$\varphi$ and $\Phi$ denote the standard normal density and cdf.
\\
{\bf Lemma 1S}.\,\,\,
\label{Le:main1} \textit{Set $\psi(a) := \int_{0}^{\infty}t^2
\Phi\left({{a}} t\right)\varphi(t)dt$,\,\,\,
$\varrho_1(a)=\frac{1}{4}+\left(\frac{1}{4\pi}+\frac{1}{8}\right)\mathbb{I}\left(a\geq1\right)$,
and
$\varrho_2(a)=\frac{1}{4}\mathbb{I}\left(a=0\right)+\left(\frac{3}{8}+\frac{a}{4\pi}\right)\mathbb{I}
\left(0<a<\frac{\pi}{2}\right)
+\frac{1}{2}\mathbb{I}\left(a\geq\frac{\pi}{2}\right)$. Then
\begin{equation*}\label{eq:le2}
\varrho_1(a) \leq \psi(a) \leq\varrho_2(a) \,\, \text{for  all}\,\,
a\geq0, \text{with   equalities for}\,\, a=0, a=1.
\end{equation*}}
\noindent {\bf   Proof}. Note that $\psi(a) =\int_{0}^{\infty}t^2
\Phi\left({{a}}t\right)\varphi(t)dt$ is increasing in $a$, and thus
for $0 \leq a < \infty$, we have $1/4 =\psi(0) \leq \psi(a) \le
\psi(\infty) = 1/2$. A simple calculation shows
that $\psi(1)=\frac{1}{4}
+\left(\frac{1}{4\pi}+\frac{1}{8}\right)$, and the lower bound follows.

The upper bound follows readily once we show that for $a>0$,\, $\psi(a) \le
\left(\frac{3}{8} +\frac{a}{4\pi}\right)$. We use the latter
inequality only for $0<a\leq\frac{\pi}{2}$  since for $a \ge
\pi/2$,\,1/2 is a better upper bound. (In fact 1/2 is a good bound since for $a>1$,
that $\psi(a) > \psi(1)=\frac{1}{4}
+\left(\frac{1}{4\pi}+\frac{1}{8}\right)\thickapprox 0.4546$.)

To show $\psi(a) \le \left(\frac{3}{8} +\frac{a}{4\pi}\right)$ for
$a>0$ we compute  Taylor's expansion around $a=1$,
\begin{align*}
&\Phi(at)=\Phi(t)+t\varphi(t)(a-1)-\frac{a^*t^3}{2}\varphi(a^*t)(a-1)^2,
\end{align*}
with $a^*$ between 1 and $a$. It follows that
\begin{align*}
&\Phi(at)\leq\Phi(t)+t\varphi(t)(a-1),\quad for\quad t\geq 0\quad
and\quad a\geq0.
\end{align*}
Therefore,
\begin{align*}
&\psi(a)=\int_{0}^{\infty}t^2
\Phi\left({{a}}t\right)\varphi(t)dt\leq \int_{0}^{\infty}t^2
\Phi\left(t\right)\varphi(t)dt+(a-1)
\int_{0}^{\infty}t^3\varphi^{2}(t)dt\\
&=\left(\frac{1}{4\pi}+\frac{3}{8}\right)+\frac{a-1}{4\pi}=
\left(\frac{3}{8} +\frac{a}{4\pi}\right) \quad for\quad all\quad
a\geq0.  \qquad \qed
\end{align*}
\\
{\bf Lemma 2S}.
\label{Le:main2}\,\,\,
\textit{Let $Z \sim N(0,1)$. Then $2\varrho_1(a)-\frac{1}{2} \leq
E\left(|Z|Z\Phi\left(aZ\right)\right) \leq
2\varrho_2(a)-\frac{1}{2}.$ Equalities hold when $a=0$ or $a=1$.}
\\
\\
\noindent {\bf   Proof}.
\begin{align*}
E\left(|Z|\Phi\left(aZ\right)Z\right)
&=\int_{-\infty}^{\infty}|t| {t} \Phi\left(at\right)\varphi(t)dt=
\int_{0}^{\infty}t^2 \Phi\left(at\right)\varphi(t)dt
-\int_{-\infty}^{0}t^2 \Phi\left(at\right)\varphi(t)dt\\
&= 2\int_{0}^{\infty}t^2
\Phi\left(at\right)\varphi(t)dt-\frac{1}{2}=2\psi(a)-\frac{1}{2} . \hspace{2.8cm} (3.1S)
\end{align*}
The result now follows   from Lemma 1S.
\qed\\
\\
\\
{\bf Lemma 3S}.
\label{Le:main3} \textit{For Model 
(1.1) with $F$ and $G$
normal, $m=2$, and $\mu=0$,
\begin{align*}
E\left(\theta_{(2)}{y}_{(2)}\right)\leq 2\sigma^2_{u}\varrho_2(a)
+\frac{\sigma^2_{e}}{\pi}\sqrt{\gamma^{*}(1-\gamma^{*})}
\end{align*}
and
\begin{align*}
E\left(\theta_{(2)}{y}_{(2)}\right)\geq 2\sigma^2_{u}\varrho_1(a)
+\frac{\sigma^2_{e}}{\pi}\sqrt{\gamma^{*}(1-\gamma^{*})},
\end{align*}
where $a=\sqrt{\frac{\gamma^{*}}{1-\gamma^{*}}}.$}
\\
\\
\noindent {\bf   Proof.} 
Kella (1986) (see also David and Nagaraja 2003) shows
that
$$ E_{{\mu}}\left(\theta_{(i)}|\bf y\right)
=\Phi\left(\bigtriangleup\right){\mu_{1}} +
\Phi\left(-\bigtriangleup\right){\mu_{2}} +(-1)^{i}\sigma \sqrt{2}
\varphi \left(\bigtriangleup\right),\eqno(3.2S)$$
 \smallskip
where
$\bigtriangleup=\gamma^{*}\frac{y_{1}-y_{2}}{\sigma\sqrt{2}}, \quad \sigma^{2}=\gamma^{*} \sigma^2_{e}
,\quad
\mu_i=\gamma^{*} y_i,\quad i=1,2.$
Therefore,
\begin{align*}
&E(\theta_{(2)}{y}_{(2)})=E({y}_{(2)}E(\theta_{(2)}|{\bf y}))
=E\left({y}_{(2)}\left(\Phi\left(\bigtriangleup\right){\mu_{1}} +
\Phi\left(-\bigtriangleup\right){\mu_{2}} +\sigma \sqrt{2} \varphi
\left(\bigtriangleup\right)\right)\right)\nonumber\\
&=\gamma^{*} E\left({y}_{(2)}
\Phi\left(\gamma^{*}\frac{y_{1}-y_{2}}{\sigma\sqrt{2}}\right)(y_1-y_2)\right)+\gamma^{*}
E\left(y_{(2)}y_2\right)+{\sigma\sqrt{2}} E\left({y}_{(2)} \varphi
\left(\gamma^{*}\frac{y_{1}-y_{2}}{\sigma\sqrt{2}}\right)\right).
\end{align*}
$$\eqno(3.3S)$$
We now  calculate  the latter three terms. For the first we use the
relation $y_{(2)}=\frac{y_1+y_2}{2}+\frac{|y_1-y_2|}{2}.$ We have
\begin{align*}
&E\left({y}_{(2)}
\Phi\left(\gamma^{*}\frac{y_{1}-y_{2}}{\sigma\sqrt{2}}\right)(y_1-y_2)\right)
=E\left(\frac{y_1+y_2}{2}\Phi\left(\gamma^{*}\frac{y_{1}-y_{2}}{\sigma\sqrt{2}}\right)(y_1-y_2)\right)\\
&+E\left(\frac{|y_1-y_2|}{2}\Phi\left(\gamma^{*}\frac{y_{1}-y_{2}}{\sigma\sqrt{2}}\right)(y_1-y_2)\right)
=E\frac{y_1+y_2}{2}E\left(\Phi\left(\gamma^{*}\frac{y_{1}-y_{2}}{\sigma\sqrt{2}}\right)(y_1-y_2)\right)\\
&+E\left(\frac{|y_1-y_2|}{2}\Phi\left(\gamma^{*}\frac{y_{1}-y_{2}}{\sigma\sqrt{2}}\right)(y_1-y_2)\right)
=E\left(\frac{|y_1-y_2|}{2}\Phi\left(\gamma^{*}\frac{y_{1}-y_{2}}{\sigma\sqrt{2}}\right)(y_1-y_2)\right).
\end{align*}
The penultimate equality follows from the fact that for iid normal
variables $y_i$, $y_{1}-y_{2}$ and $y_{1}+y_{2}$ are independent,
and the last equality holds because for $\mu=0$ we have $E(y_i)=0$.
The substitution
$Z=\frac{y_1-y_2}{[2(\sigma^2_u+\sigma^2_e)]^{1/2}}$ and standard
calculations show that the last term above equals
\begin{align*}
&\frac{\sigma^2_{u}}{\gamma^{*}}E\left(|Z|\Phi\left(\sqrt{\frac{\gamma^{*}}{1-\gamma^{*}}}Z\right)Z\right),
\end{align*}
where $Z$ is a  standard normal random variable.\\

Let $a=\sqrt{\frac{\gamma^{*}}{1-\gamma^{*}}}$. Using
(3.1S) we obtain
\begin{align*}
&E\left({y}_{(2)}
\Phi\left(\gamma^{*}\frac{y_{1}-y_{2}}{\sigma\sqrt{2}}\right)(y_1-y_2)\right)=
\frac{\sigma^2_{u}}{\gamma^{*}}E\left(|Z|\Phi\left(a\,Z\right)Z\right)
=\frac{\sigma^2_{u}}{\gamma^{*}}(2\psi(a)-1/2).
\end{align*}

To calculate the second term of
(3.3S) we use a result
from Siegel (1993), see also Rinott and Samuel-Cahn (1994). It
yields the second equality below, while the others are
straightforward:
$$E\left(y_{(2)}y_2\right)=Cov(y_2,y_{(2)})=Cov(y_2,y_2)P(y_2=y_{(2)})+Cov(y_2,y_1)P(y_2=y_{(1)})
= \frac{\sigma^2_{u}}{2\gamma^{*}}.$$
\ignore{\begin{align*}
&y_{(2)}=\frac{y_1+y_2}{2}+\frac{|y_1-y_2|}{2}\Longrightarrow
E(y_{(2)})=\mu+\frac{1}{2}E|y_1-y_2|=\mu+\frac{\sqrt{2(\sigma^2_{u}+\sigma^2_{e}})}{2}E|Z|,
\end{align*}
where $Z$ is a  standard normal random variable.\\
Hence,
\begin{align*}
&E(y_{(2)})=\mu+\frac{\sqrt{2(\sigma^2_{u}+\sigma^2_{e}})}{2}\sqrt{\frac{2}{\pi}}=\mu+\frac{\sqrt{\sigma^2_{u}+\sigma^2_{e}}}{\pi}.
\end{align*}
} 
The third part of
(3.3S) is computed like the second
part above to give
\begin{align*}
E\left({y}_{(2)} \varphi
\left(\gamma^{*}\frac{y_{1}-y_{2}}{\sqrt{2}}\right)\right)=
\sqrt{\frac{\sigma^2_{u}}{2\gamma^{*}}}E\left(|Z| \varphi
\left(a\,Z\right)\right).
\end{align*}
The latter expectation becomes
\begin{align*}
 \int_{-\infty}^{\infty}|t|\varphi
\left(a\,t\right)\varphi(t)dt &=2\int_{0}^{\infty}t\,\varphi
\left(a\,t\right)\varphi(t)dt=\frac{1-\gamma^{*}}{\pi}.
\end{align*}
Combining these results, we get
$$E\left(\theta_{(2)}{y}_{(2)}\right)=
2\sigma^2_{u}\psi(a)
+\frac{\sigma^2_{e}}{\pi}\sqrt{\gamma^{*}(1-\gamma^{*})}. \eqno (3.4S)
$$
From
(3.4S) and Lemma 1S, Lemma 3S
follows readily.\qed
\\
\\
\noindent {\bf  Proof of Theorem 5}.
It is easy to see that we can assume $\mu=0$ without loss of generality.
We  use the calculations of  Theorem 3.
Lemma 3S
is used instead of Lemma 1
for a better
upper bound of
$E\left(\theta_{(1)}y_{(1)}+\theta_{(2)}y_{(2)}\right)$ for the
normal case and $m=2$.

Below we use the notation of Lemma 1. By symmetry
$E(\theta_{(1)}{y}_{(1)})=E(\theta_{(2)}{y}_{(2)})$. Therefore, by
Lemma 3S with
$a=\sqrt{\frac{\gamma^{*}}{1-\gamma^{*}}}$\,, we have
$E(\theta_{(1)}{y}_{(1)}+\theta_{(2)}{y}_{(2)})\leq 4\sigma^2_{u}
\varrho_2(a)
+2\frac{\sigma^2_{e}}{\pi}\sqrt{\gamma^{*}(1-\gamma^{*})}.$
By (7.2)
 and the above inequality we obtain
\begin{align*}
&E\sum_{i=1}^2\left({y}_{(i)}-\theta_{(i)})({y}_{(i)}-\overline{y}\right)=
2(\sigma_u^2+\sigma^2_{e})-\sigma^2_{e}-E\sum_{i=1}^2{\theta_{(i)}}{y}_{(i)}\\
&\geq 2(\sigma_u^2+\sigma^2_{e})-\sigma^2_{e}-4\sigma^2_{u}
\varrho_2(a)
-2\frac{\sigma^2_{e}}{\pi}\sqrt{\gamma^{*}(1-\gamma^{*})}\\
&=2\sigma_u^2-4\sigma^2_{u}
\varrho_2(a)+\sigma^2_{e}\left(1-\frac{2}{\pi}\sqrt{\gamma^{*}
(1-\gamma^{*})}\right) :=\kappa(\gamma^*).
\end{align*}
Recall from the proof of Theorem 3 the notation
\begin{align*}
&D(\gamma):=E\{L({\boldsymbol{\widehat{\theta}}{\,\,^{[2]}_{(\,)}}(\gamma),
\boldsymbol{\theta}_{(\,)})}\}- E\{L({\boldsymbol{\widehat{\theta}}{\,\,^{[1]}_{(\,)}},
\boldsymbol{\theta}_{(\,)})}\}.
\end{align*}
In order to prove part 1 of Theorem 5, we have to show that its conditions imply $D(\gamma) \le 0$.

By (7.3)
for $m=2$,
\begin{align*}
&D(\gamma)=(1-\gamma)^{2} (\sigma_u^2+\sigma^2_{e})
-2(1-\gamma)E\sum_{i=1}^2({y}_{(i)}-\theta_{(i)})({y}_{(i)}-\overline{y})\\
&\leq (1-\gamma)^{2}
(\sigma_u^2+\sigma^2_{e})-2(1-\gamma)\kappa(\gamma^*)
=(1-\gamma)[(1-\gamma)
(\sigma_u^2+\sigma^2_{e})-2\kappa(\gamma^*)].
\end{align*}
We assume $0 \le \gamma \le 1$ and therefore $D(\gamma)\leq 0$
provided
 $\gamma\geq1-2\frac{\kappa(\gamma^*)}{\sigma_u^2+\sigma^2_{e}}=:\omega(\gamma^*)$.

For $\gamma^{*}=0 \,\,(a=0)$, $\omega(\gamma^{*})=-1$ and clearly $D(\gamma) \le 0$ for all $\gamma$.

\ignore{
Using Descartes' rule of signs and some calculations, it can be
shown (Malinovsky 2008) that the function $\omega$ has a single zero
on the positive real line at some $c>0$ and  $\gamma^*<c$ implies
$\omega(\gamma^*)<0$.  By numerical calculation we obtain
$c=0.4119$.}

Next we show that in the range
$\displaystyle 0<\gamma^*<\frac{\pi^2}{\pi^2+4} \approx 0.71 \,\left(0<a<\frac{\pi}{2}\right)$
the function $\omega(\gamma^*)$ has a single zero at $c \approx 0.4119$, and $\omega(\gamma^*)<0$
for $\gamma^* <c$. This implies that $\gamma>\omega(\gamma^*)$
and therefore $D(\gamma)<0$.

In this range of $\gamma^{*}$,\,
$\omega(\gamma^{*})=1+4\gamma^{*}\left( \frac{a}{2\pi}-\frac{1}{4}\right)-2
(1-\gamma^{*})\left(1-\frac{2}{\pi}\sqrt{\gamma^{*}
(1-\gamma^{*})}\right)$.
Substituting $\displaystyle \gamma^*=\frac{a^2}{1+a^2}$ we get
$\omega(\gamma^{*})=1+\frac{1}{1+a^2}\left(\frac{2}{\pi}a^3-a^2-2+\frac{4}{\pi}\frac{a}{1+a^2}\right)$.
The function $\omega(\gamma^*)$ has the same zeros as the function
$ P(a):=\frac{\pi}{2}(1+a^2)^2\omega(\gamma^*)$
and straightforward calculations show that
$P(a)=a^5+a^3-\frac{\pi}{2}a^2+2a-\frac{\pi}{2}$, and that this function is increasing in $a$ and therefore
in $\gamma^{*}$.
By numerical calculation we obtain that it vanishes at
$c \approx 0.4119$.


The second part of Theorem 5
is proved by showing that   $\gamma^*\geq
\omega(\gamma^{*})$ and therefore $ 1\geq\gamma\geq\gamma^{*}$
implies $\gamma\geq \omega(\gamma^{*})$.


In the range
$\displaystyle 0<\gamma^*<\frac{\pi^2}{\pi^2+4} \approx 0.71 \,\left(0<a<\frac{\pi}{2}\right)$,
$\omega(\gamma^{*})=1+4\gamma^{*}\left( \frac{a}{2\pi}-\frac{1}{4}\right)-2
(1-\gamma^{*})\left(1-\frac{2}{\pi}\sqrt{\gamma^{*}
(1-\gamma^{*})}\right)\leq
1-2
(1-\gamma^{*})\frac{\pi-1}{\pi}$\,. Therefore, $\omega(\gamma^{*})-\gamma^{*}\leq\frac{2-\pi}{\pi}(1-\gamma^{*})<0$.

In the range
$\displaystyle \gamma^*\geq\frac{\pi^2}{\pi^2+4}$\,,
$\omega(\gamma^{*})=1-2
(1-\gamma^{*})\left(1-\frac{2}{\pi}\sqrt{\gamma^{*}
(1-\gamma^{*})}\right)\leq
1-2
(1-\gamma^{*})\frac{\pi-1}{\pi}\, .$
Therefore, $\omega(\gamma^{*})-\gamma^{*}\leq\frac{2-\pi}{\pi}(1-\gamma^{*})<0$.

For the proof the last part we use the same calculation as in Theorem 4
with  m=2 to
obtain
\begin{align*}
&\partial E\{L(\boldsymbol{\widehat{\theta}}\,^{[2]}_{(\,)}(\gamma),
\boldsymbol{\theta}_{(\,)})\}/\partial{\gamma}=0  \quad\text{if and only if}\quad
\gamma=1-\frac{E\sum_{i=1}^2\left({y}_{(i)}-\theta_{(i)})({y}_{(i)}-\overline{y}\right)}{(\sigma_u^2+\sigma^2_{e})}.
\end{align*}
By (7.2)
 we have
\begin{align*}
&E\sum_{i=1}^2\left({y}_{(i)}-\theta_{(i)})({y}_{(i)}-\overline{y}\right)=
2(\sigma_u^2+\sigma^2_{e})-\sigma^2_{e}-E\sum_{i=1}^2{\theta_{(i)}}{y}_{(i)}.
\end{align*}
By (3.4S)
we have $E\sum_{i=1}^2{\theta_{(i)}}{y}_{(i)}=
4\sigma^2_{u}\psi(a)
+2\frac{\sigma^2_{e}}{\pi}\sqrt{\gamma^{*}(1-\gamma^{*})}$.\,\,
Hence,
\begin{align*}
&E\sum_{i=1}^2\left({y}_{(i)}-\theta_{(i)})({y}_{(i)}-\overline{y}\right)=
2\sigma^2_u\left(1-2\psi(a)\right)+\sigma^2_e
\left(1-\frac{2}{\pi}\sqrt{\gamma^{*}(1-\gamma^{*})}\right).
\end{align*}
Finally, using the convexity of
$E\{L({\widehat{\theta}}{\,^{[2]}}(\gamma), \theta)\}$, the optimal
$\gamma$ is
\begin{align*}
\gamma^{o}=\gamma^*\left(4\psi(a)-1\right)+(1-\gamma^*)\frac{2}{\pi}\sqrt{\gamma^{*}(1-\gamma^{*})}.
\qquad \qquad\qed
\end{align*}
\subsection{Proof of Theorem 6}
Note that
${\widehat{\theta}_{(i)}}^{\,\,{[2]}}(\gamma)=(1-\gamma)\overline{y}+\gamma
g_{i}(y)$, and  from (3.2S)
${\widehat{\theta}_{(i)}}^{\,\,{[3]}}=E_{\widehat{\mu}}\left(\theta_{(i)}|y\right)=(1-\gamma^{*})\overline{y}+\gamma^{*}
f_{i}(y)$, where $f_i(y)$ and $g_i(y)$ are functions of  $y=(y_1,y_2)$ defined for $i=1,2$
by
\begin{align*}
&f_i\equiv f_{i}(y)=
(-1)^{i}\left(\Phi\left(\bigtriangleup\right)(y_{1}-y_{2})
+\frac{\sigma}{\gamma^*} \sqrt{2} \varphi
\left(\bigtriangleup\right)\right)+y_{i},\,\,\,g_i\equiv g_{i}(y)=y_{(i)},\\
&\bigtriangleup=\gamma^*\frac{y_{1}-y_{2}}{\sigma\sqrt{2}},\quad
\sigma^{2}=\gamma^{*} \sigma^2_e.
\end{align*}
We have
\begin{align*}
&E\left(({\widehat{\theta}_{(i)}}^{\,\,{[3]}}-\theta_{(i)})^2|
y\right)=Var\left(\theta_{(i)}|y\right)+\left(E\left((\theta_{(i)}-{\widehat{\theta}_{(i)}}^{\,\,{[3]}})|
y\right)\right)^{2}\\
&=Var\left(\theta_{(i)}|y\right)+\left((1-\gamma^*)\left(\mu-\overline{y}\right)\right)^{2}=
Var\left(\theta_{(i)}|y\right)+\left((1-\gamma^*)\overline{y}\right)^{2},
\end{align*}
where the last equality holds
because for m=2 we can assume that  $\mu=0$  without loss of generality. In the same way,
\begin{align*}
&E\left(({\widehat{\theta}_{(i)}}^{\,\,{[2]}}(\gamma)-\theta_{(i)})^2|
y\right)=Var\left(\theta_{(i)}|y\right)+\left(E\left((\theta_{(i)}-{\widehat{\theta}_{(i)}}^{\,\,{[2]}}(\gamma))|
y\right)\right)^{2}\\
&=Var\left(\theta_{(i)}|y\right)+\left((1-\gamma^{*})\mu+\gamma^{*}
f_{i}-(1-\gamma)\overline{y}-\gamma g_{i}\right)^{2}
=Var\left(\theta_{(i)}|y\right)+\left(\gamma^{*}
f_{i}-(1-\gamma)\overline{y}-\gamma g_{i}\right)^{2}.
\end{align*}
Therefore,
\begin{align*}
&d(\gamma):=E\{L({\boldsymbol{\widehat{\theta}}{\,\,^{[3]}_{(\,)}},\boldsymbol{\theta}_{(\,)})}\}-
E\{L({\boldsymbol{\widehat{\theta}}{\,\,^{[2]}_{(\,)}}(\gamma),\boldsymbol{\theta}_{(\,)})}\}\\
&=\sum_{i=1}^2
E\left\{E\left(({\widehat{\theta}_{(i)}}^{\,\,{[3]}}-\theta_{(i)})^2|
y\right)\right\} -\sum_{i=1}^2 E\left\{  E\left(({\widehat{\theta}_{(i)}}^{\,\,{[2]}}(\gamma)-\theta_{(i)})^2| y\right)  \right\} \\
&=2E\left((1-\gamma^*)\overline{y}\right)^{2}- E\left(\gamma^{*}
f_{1}-\gamma g_{1}-(1-\gamma)\overline{y}\right)^{2}
-E\left(\gamma^{*}
f_{2}-\gamma g_{2}-(1-\gamma)\overline{y}\right)^{2}\\
&=2\left((1-\gamma^*)^2-(1-\gamma)^2\right)E\left(\overline{y}\,^2\right)
-E\left(\gamma^{*} f_{1}-\gamma g_{1}\right)^{2}-E\left(\gamma^{*}
f_{2}-\gamma g_{2}\right)^{2}\\
&+2(1-\gamma)E\big[\left((\gamma^{*}( f_{1}+f_{2})-\gamma
(g_{1}+g_{2})\right)(\overline{y})\big].
\end{align*}
From the definitions of $f_{i}$ and $g_{i}$ it follows that
$f_{1}+f_{2}-g_{1}-g_{2}\equiv 0$, and the last term vanishes. It
is now easy to see that $d(\gamma^*) \le 0$. \qed
\end{document}